\documentclass{emulateapj}

\usepackage{natbib}
\usepackage{subfigure}
\usepackage{graphicx}
\usepackage{lscape}
\usepackage{rotating}
\usepackage{longtable}

\newcommand{\kms}{km~s$^{-1}$}
\newcommand{\sdom}{$\sigma_{m}$}
\newcommand{\sd}{$\sigma$}
\newcommand{\onesigma}{$1\sigma$}
\newcommand{\threesigma}{$3\sigma$}

\newcommand{\Halpha}{H$\alpha$}
\newcommand{\Hbeta}{H$\beta$}
\newcommand{\NIIHalpha}{$\log_{10}$($[$\ion{N}{2}$]$/H{$\alpha$})}
\newcommand{\OIIIHbeta}{$\log_{10}$($[$\ion{O}{3}$]$/H{$\beta$})}
\newcommand{\taus}{$\tau_s$}
\newcommand{\chisq}{$\chi^{2}$}
\newcommand{\lb}{$\left<\right.$}
\newcommand{\rb}{$\left.\right>$}
\newcommand{\fvar}{F_{\mathrm{var}}}
\newcommand{\fvarm}{$F_{\mathrm{var}}$}
\newcommand{\dfoverfm}{$\Delta{f}/{f}$}
\newcommand{\structf}{{\it SF}}
\newcommand{\radeccrossmatcharcsec}{1\arcsec}
\newcommand{\radectolinarcsec}{1}
\newcommand{\redshifttol}{0.01}
\newcommand{\qsoredshiftmin}{0.08}
\newcommand{\qsoredshiftmax}{3.51}
\newcommand{\qsoredshiftavg}{1.16}
\newcommand{\stellarabs}{1.3~\AA}
\newcommand{\sdssspectrophotouncertainty}{$7\%$ at observed frame wavelength $3800$~\AA}
\newcommand{\oiiiavgsdssdrsix}{a~few~\AA}
\newcommand{\sdssspecclassfwhm}{1000~\kms}
\newcommand{\sdssspecclassstrongline}{10~\AA}
\newcommand{\sdssspecclassstronglinesigma}{$> 3\sigma$}
\newcommand{\obswaverange}{$3900-8900$~\AA}
\newcommand{\mjdrange}{$\sim{3 -  700}$~days}
\newcommand{\specimprovement}{two}
\newcommand{\specimprovementStar}{three}
\newcommand{\npairtotalStar}{9,100}
\newcommand{\npairtotalQSO}{3,205}
\newcommand{\npairtotal}{23,330}
\newcommand{\npair}{18,435}
\newcommand{\timebaseline}{$\sim{700}$~days}
\newcommand{\platesdropped}{$1664,1905,1907,2075$}
\newcommand{\nplate}{60}
\newcommand{\npairminCali}{10}

\newcommand{\SNmin}{$\geq 10$}
\newcommand{\npairECLASSEF}{$\leq~3$}
\newcommand{\fvarMeanDropLowSN}{$0.010$}
\newcommand{\fvarSDDropLowSN}{$0.003$}
\newcommand{\varinSigma}{$> 3\sigma$}
\newcommand{\variMaxSigma}{$4\sigma$}
\newcommand{\fvarfullsampleAGN}{$0.9\%$}
\newcommand{\fvarfullsampleSF}{$1.2\%$}
\newcommand{\dgSEYFERT}{$0.07  \pm  0.26$~mag}

\newcommand{\dgfluxSEYFERTUpperLimit}{$\sim{30}$\%}

\newcommand{\ewMin}{$1$~\AA}
\newcommand{\nsigmaMin}{2$\sigma$}
\newcommand{\dlineratioxSEYFERT}{$0.02  \pm  0.03$~dex}
\newcommand{\dlineratioySEYFERT}{$0.06  \pm  0.08$~dex}
\newcommand{\dtaurest}{$\Delta\tau_{\mathrm rest}$}
\newcommand{\repeatabilitymin}{0.010~or~1.0\%}

\newcommand{\contvarirangegdiff}{$0.07-0.10$~mag}
\newcommand{\tausGALAXY}{$\sim{10}$~years}
\newcommand{\tausUncertaintyGALAXY}{$\sim{2}$~years}
\newcommand{\tausChisqGALAXY}{$20$}
\newcommand{\tausQSO}{$0.4\pm   1.0$~years}
\newcommand{\dtauRestRangeGALAXY}{$0-500$}
\newcommand{\dtauRestRangeQSO}{$0-300$}
\newcommand{\dtauRestRangeStar}{$0-710$}
\newcommand{\matchQSOSouthStripe}{$300$}
\newcommand{\matchQSOSouthStripeVari}{$10\%$}

\begin{document}

\title{Probing Spectroscopic Variability of Galaxies \& Narrow-Line
Active Galactic Nuclei in the Sloan Digital Sky Survey}

\author{C.~W.~Yip\altaffilmark{1},      A.~J.~Connolly\altaffilmark{2},
D.~E.~Vanden~Berk\altaffilmark{3},                         R.~Scranton\altaffilmark{4},
S.~Krughoff\altaffilmark{2},              A.~S.~Szalay\altaffilmark{1},
L.~Dobos\altaffilmark{5}, C.~Tremonti\altaffilmark{6},  M.~Taghizadeh-Popp\altaffilmark{1},
T.~Budav\'ari\altaffilmark{1},               I.~Csabai\altaffilmark{5},
R.~F.~G.~Wyse\altaffilmark{1}, \v{Z}.~Ivezi\'{c}\altaffilmark{2}}
 
\altaffiltext{1}{Department  of  Physics   and  Astronomy,  The  Johns
Hopkins University, 3701 San Martin Drive, Baltimore, MD 21218, USA.}  
\altaffiltext{2}{Astronomy   Department,   University   of
Washington, WA 98195, USA.}  
\altaffiltext{3}{Saint Vincent College, 300 Fraser Purchase Road,
  Latrobe, PA 15650, USA.}
\altaffiltext{4}{Department of Physics and Astronomy, University of
  Pittsburgh, 3941 O'Hara Street, Pittsburgh, Pennsylvania 15260, USA}
\altaffiltext{5}{Department of Physics of
Complex Systems, E\"{o}tv\"{o}s  Lor\'and University, H-1117 Budapest,
Hungary.}
\altaffiltext{6}{Steward Observatory, University of Arizona, 933
  N. Cherry Avenue, Tucson AZ 85721, USA.}

\begin{abstract}
  Under  the   unified  model  for  active   galactic  nuclei  (AGNs),
narrow-line (Type 2)  AGNs are, in fact, broad-line  (Type 1) AGNs but
each  with a  heavily  obscured accretion  disk.   We would  therefore
expect the optical  continuum emission from Type 2  AGN to be composed
mainly of stellar light and  non-variable on the time-scales of months
to  years.  In  this work  we probe  the spectroscopic  variability of
galaxies and narrow-line AGNs using  the multi-epoch data in the Sloan
Digital Sky Survey (SDSS) Data Release 6. The sample contains \npair \
sources  for which  there  exist pairs  of spectroscopic  observations
(with  a  maximum separation  in  time  of  \timebaseline) covering  a
wavelength range of \obswaverange.  To obtain a reliable repeatability
measurement  between  each spectral  pair,  we  consider  a number  of
techniques for spectrophotometric calibration resulting in an improved
spectrophotometric calibration of  a factor of \specimprovement.  From
these data we find  no obvious continuum and emission-line variability
in the narrow-line AGNs on average -- the spectroscopic variability of
the  continuum  is  \dgSEYFERT  \   in  the  $g$  band  and,  for  the
emission-line ratios  \NIIHalpha \ and \OIIIHbeta,  the variability is
\dlineratioxSEYFERT \ and \dlineratioySEYFERT, respectively.  From the
continuum variability measurement  we set an upper limit  on the ratio
between the flux of  varying spectral component, presumably related to
AGN    activities,    and    that     of    host    galaxy    to    be
\dgfluxSEYFERTUpperLimit.  We  provide the corresponding  upper limits
for  other spectral  classes, including  those from  the  BPT diagram,
eClass galaxy classification, stars and quasars.
\end{abstract}


\section{Introduction}

Active galactic nuclei (AGNs) are  found to vary in the time-domain at
many  frequencies.   The  underlying  physical mechanism  driving  the
variability      seen     in      AGN      spectra     is      unknown
\citep[e.g.,][]{2001sac..conf....3P,2003MmSAI..74.1004B}.       Several
scenarios  have  been proposed  including  accretion disk  instability
\citep{1998ApJ...504..671K},            supernova            explosion
\citep{1997MNRAS.286..271A}, microlensing by  stars in the intervening
galaxy \citep{1993Natur.366..242H}, and varying ionizing source or/and
varying   optical  depth   of  the   material  in   the   vicinity  of
light-generating                                                regions
\citep[e.g.,][]{1976ApJ...210L.117T,1989ApJ...340..190G}.  At present,
models  involving  non-thermal  emission   from  the  jets  or  events
occurring  on the  AGN  accretion disk  are  most favored  \citep[see,
e.g.,][for  a   summary  of  variability  models  in   AGNs  by  radio
loudness]{2004MNRAS.350..175S}.

Most  variability  studies,  however,  have  been focused  on  Type  1
(broad-line) AGNs.   While the emission lines of  Type 2 (narrow-line)
AGNs were  found to vary  in the X-rays  on time-scales from  hours to
years
\citep{1998MNRAS.298..824G,2003HEAD....7.2110M,2007A&A...473..105M},
their optical continua and  emission lines are generally considered to
be non-variable.  For the  emissions from the narrow-line regions, the
spatially     low-density     region     (electron    density     $n_e
\sim{1,000}$~cm$^{-3}$)   results   in   a  long   ($\sim{100}$~years)
recombination  time\footnote{The   recombination  time  $\tau_{e}$  is
referred to that of the hydrogen  atom, so that $\tau_{e} = 1 / (n_{e}
\,  \alpha_{A})$, where $n_{e}$  is the  electron density.   The total
recombination coefficient of hydrogen, $\alpha_{A}$, is equal to $4.18
\times 10^{-13}$~cm$^{3}$~s$^{-1}$ at an electronic temperature $T_{e}
=                10^{4}$~K               \citep{1989agna.book.....O}.}
\citep{1995PASP..107..579P},  exceeding  the  duration  of  a  typical
observation.  For the optical  continuum, it is commonly considered to
be  composed  mainly  of  the   stellar  light,  in  accord  with  the
expectation         from         the         unification         model
\citep{1993ARA&A..31..473A,1995PASP..107..803U}    in   which   Type~2
objects  are  obscured  by  a  dusty torus  along  the  line-of-sight.
Dominated by the  stellar light, the optical continuum  in Type 2 AGNs
should therefore  be temporally  non-variable.  There is,  however, no
strong observational evidence in  the time-domain to either support or
dispute these ideas.

From a 4-year  monitoring of the broad-band optical  variability of 35
Type~1    and   2   Seyfert    galaxies,   \citet{1992MNRAS.257..659W}
demonstrated  that  most  galaxies  in  their  sample  were  variable.
Another promising work to probe  Type 2 AGN variability in the optical
is  the intra-night  variability  of Seyfert  2  galaxies observed  by
\citet{2001Ap&SS.275..209J}, where two out of the three objects varied
by  $\sim{0.25}$~mag.   Further,  \citet{2008AJ....135.2048T} found  a
Type~1.9 Seyfert  galaxy \citep{1989agna.book.....O} to  change into a
Type~2 Seyfert over a few  years, and proposed the underlying cause to
be a varying ionizing continuum.

The UV-optical  variability of  broad-line AGNs is  better established
with  an  amplitude   of  $\sim{10\%}$  \citep[e.g.,  see  discussions
in][]{2004ApJ...601..692V}.  Wavelength  dependence of QSO variability
has   been    studied   by   \citet{2005ApJ...633..638W},    and   the
\ion{C}{4}-emission  dependence by \citet{2006ApJ...641...78W}  in the
Sloan  Digital  Sky  Survey  \cite[SDSS,][]{2000AJ....120.1579Y}.   In
these  studies   the  authors  have  developed  a   method  to  remove
wavelength-dependent systematics which may  exist in the spectra, from
a  spectroscopic  plate\footnote{A  spectroscopic  plate in  the  SDSS
contains a set  of spectra which are observed  simultaneously.} at one
epoch of observation  to the same plate at  another epoch.  Correcting
for this on  a plate-to-plate basis they were  able to select variable
QSOs  and to  demonstrate  a power-law-like  QSO difference  spectrum;
which in turn can be explained  by the change in the accretion rate in
thermal  accretion disk \citep{2006ApJ...642...87P}.   QSO photometric
variability     in     the     SDSS     has    been     studied     by
\citet{2004ApJ...601..692V} for its  dependency on physical parameters
such    as   redshift,   luminosity    and   radio    properties;   by
\citet{2003AJ....126.1217D} in  which the  authors used both  the SDSS
and the historical observations (up to $\sim{50}$~years) and found the
time-scale  of   the  variability   to  be  $\sim{2}$~years;   and  by
\citet{2004IAUS..222..525I}  and \citet{2006AJ....131.2801S}  in which
the authors  used the SDSS  repeated imagings in combination  with the
Palomar  Observatory Sky  Survey (POSS)  data to  probe long  term QSO
variability (up  to $\sim{50}$~years), and constrained  its time scale
to be $\sim{1}$~year in the restframe.

Normal ``inactive''  galaxies (for our purpose defined  to be galaxies
that  are  not  Type~1  or  Type~2 AGNs,  nor  low-ionization  nuclear
emission  regions --  \citep[LINERS,][]{1980A&A....87..152H})  are not
observed to  be variable; and are  generally not expected  to be.  For
the continua of galaxies, the  time scale of observation is negligible
compared with that of  stellar evolution (for solar-mass main sequence
stars, lifetime $\approx{10^{10}}$~years).  For emission-line regions,
that  arise  from  low-density  environments and  \ion{H}{2}  regions,
similar arguments can  be made to those for  narrow-line regions, with
the  understanding that  the electron  density is  likely to  be lower
\citep[$n_{e} \sim{1 - 1,000}$~cm$^{-3}$,][]{1978ppim.book.....S}.

The goal of this work  is, therefore, to probe spectral variability of
galaxies and narrow-line AGNs using the SDSS multi-epoch observations.
The SDSS  data have the advantage  of large sample size  with the same
spectrophotometric reduction, which is critical to our work because we
expect any variability present in  galaxies to be small (compared with
the  amplitude  of  the  QSO  UV-optical  variability,  for  example).
Several techniques  are used to refine  the spectroscopic calibration,
including those by  \citet{2005ApJ...633..638W}.  The resultant galaxy
variability  measurement is  compared  with that  of  stars and  QSOs,
independently analyzed in this work.

In   \S\ref{section:data}   we   discuss   the   samples   used.    In
\S\ref{section:fvar}   we  present   the   variability  measure.    In
\S\ref{section:recalibration}  we  describe   the  refinement  of  the
spectroscopic  calibration.  In  \S\ref{section:vari}  we present  the
galaxy continuum variability, and the emission line variability of the
variable     candidates     based     on    the     continuum.      In
\S\ref{section:comparestarqso}  we   present  results  on  variability
analyses    on    stars   and    QSOs    as    sanity   checks.     In
\S\ref{section:conclusion} we summarize the results and discuss future
work.

\section{Samples} \label{section:data}

As part  of the SDSS \citep{2000AJ....120.1579Y}  spectra are observed
with fibers of 3\arcsec \, diameter \citep[corresponding to 0.18~mm at
the focal plane for the 2.5~m, f/5 telescope,][]{2006AJ....131.2332G}.
The  spectral resolution  $R$ is  $\sim{1800}$ in  the  observed frame
$3800-9200$~\AA. All  sources are  selected (or ``targeted'')  from an
initial  imaging  survey  using   the  SDSS  camera  as  described  in
\citet{1998AJ....116.3040G}  with   the  filter  response   curves  as
described  in   \cite{1996AJ....111.1748F},  and  using   the  imaging
processing  pipeline of  \cite{2001ASPC..238..269L}.   The astrometric
calibration   is   described   in   \cite{2003AJ....125.1559P}.    The
photometric    system    and     calibration    are    described    in
\citet{1996AJ....111.1748F},               \citet{2001AJ....122.2129H},
\citet{2002AJ....123.2121S}      and      \citet{2004AN....325..583I},
\citet{2006AN....327..821T}.    The   targeting   strategy   for   the
multi-object spectrograph is described in \citet{2003AJ....125.2276B}.

We    select    our    samples     from    the    Data    Release    6
\citep[DR6,][]{2008ApJS..175..297A} of the SDSS.  The galaxy sample is
constructed     from      the     SDSS     Main      Galaxy     sample
\citep{2002AJ....124.1810S},  by  selecting  spectra  which  are  both
spectroscopically  classified as  galaxies, and  rejecting  spectra in
which the  redshift measurements  failed or have  not been  made.  The
spectral classification (star, galaxy,  QSO) is highly confident, with
98\% of  all the DR6 spectra having  consistent classification between
the  spectro1d  pipeline  \citep{2002SPIE.4847..452S} and  the  specBS
pipeline  (Schlegel et~al.   in prep.).   From this  sample,  pairs of
spectral observations  are identified by  matching sources in  both RA
and DEC to within  \radectolinarcsec\arcsec, and in redshift to within
\redshifttol.  Spectra  at both epochs  are required to have  the same
plate  number but different  date of  observations, given  in Modified
Julian Date (MJD) by the SDSS.  The selection results in \npairtotal \
pairs.  The  subsequent sample selections  (mainly for the  purpose of
increasing the signal-to-noise, S/N, of the spectra) will be described
in     \S\ref{section:recalibration}     and     \S\ref{section:vari}.
Table~\ref{tab:plate} lists the  relevant spectroscopic plates and the
corresponding number of galaxy spectra.  The sample covers \mjdrange \
in the observed frame.

The  spectroscopically  classified  galaxies  selected  above  do  not
contain any strong  (equivalent width, EW $>$\sdssspecclassstrongline,
and  \sdssspecclassstronglinesigma \  detection in  the height  of the
line)    broad   line    (full   width    at   half    maximum,   FWHM
$>$\sdssspecclassfwhm)  \citep{2002SPIE.4847..452S}.   The  cutoff  is
below the  traditional line-width selection  for broad lines  (Type 1)
galaxies,  where  the  width  of  permitted line(s)  $>  2000$~\kms  \
\citep[e.g.,][]{2008RMxAC..32...86K}.  The  sample therefore comprises
galaxies  of different  eClass types  \citep{1995AJ....110.1071C}, and
narrow-line  AGNs.  A  potential  contamination to  our galaxy  sample
would  be the  narrow-line Seyfert~1  galaxies  \cite[][and references
therein]{1985ApJ...297..166O}, expected to offer  a direct view of the
active nuclei  \citep{1993ARA&A..31..473A,1995PASP..107..803U} and can
be            variable            in            the            optical
\citep{1996A&A...314..419G,2000NewAR..44..491P}.     Each   of   these
spectra shows a  broad component in Balmer emission  line with FWHM $<
2000$~\kms,  and  relatively  weak  $[$\ion{O}{3}$]\lambda$5008.   The
reasons behind  this potential  contamination are twofold  -- firstly,
the spectro1d pipeline  fits to each line a  single Gaussian, that may
underestimate the line  width of the broad component in  the case of a
composite  line  which  is  made  up  of both  the  broad  and  narrow
components,  in particular  when  the broad  one  is relatively  weak.
Secondly, the FWHM of the broad component of both \Halpha \ and \Hbeta
\  of a  narrow-line Seyfert~1  can  extend down  to $\sim500$~\kms  \
\citep[Fig.~4  of][]{2006ApJS..166..128Z}, falling  within  the galaxy
spectral classification  in the SDSS.   However, we expect  the number
contribution of these objects to the whole DR6 galaxy+QSO sample to be
small, $\sim0.5$\%, with  reference to \citet{2006ApJS..166..128Z} who
used an upper limit of $2200$~\kms \ in the broad component of \Halpha
\ or  \Hbeta \  emissions to search  for narrow-line Seyfert~1  in the
SDSS.  Similarly, the Seyfert~1.8 and~1.9 \citep{1989agna.book.....O},
both of  which show  strong narrow components  (relative to  the broad
ones, if  available) in \Halpha \  and \Hbeta, can be  variable in the
optical \citep{1989ApJ...340..190G}, and may  be present in our galaxy
sample as  well.  We note that  the current analysis  does not provide
further  separation of  those  from our  emission-line galaxies,  that
calls  for  fitting of  double  Gaussian  to  each Balmer  line,  for
example.  The  number contribution of  these types to the  full galaxy
sample     is      expected     to     be      small,     $\sim0.02$\%
\citep{2008ApJ...679...86W}.   In  this  work,  the  fraction  of  the
classified Seyfert~2  + starforming +  composite galaxies to  the full
sample of our galaxy spectral pairs is $\sim14$\%; and that of all the
eClass        types,         $\sim22$\%        (calculated        from
Table~\ref{tab:fvarFullType}).

When constructing the stellar and  QSO samples, the spectral pairs are
selected  similarly to  the case  of  galaxies, except  that they  are
spectroscopically  classified  as  star  and QSO,  respectively.   The
targeting selection for spectroscopic  observation of QSOs in the SDSS
is  described  in   \citet{2002AJ....123.2945R}.   Each  QSO  spectrum
classified  \citep{2002SPIE.4847..452S} has at  least one  strong line
(see    above    for   the    galaxy    classification)   with    FWHM
$>$\sdssspecclassfwhm,  or/and  a  Lyman  alpha  forest.   Unlike  the
selection of the galaxies and the QSOs, no criterion is imposed on the
quality of the redshift when  selecting the stars.  The upper bound of
their   distances   is   $cz    =   450$~\kms,   or   $z   =   0.0015$
\citep{2002SPIE.4847..452S}.   The samples  contain  \npairtotalStar \
stellar pairs and \npairtotalQSO  \ QSO pairs (\qsoredshiftmin \ $<z<$
\qsoredshiftmax), respectively.

The  spectra are  de-reddened  against Galactic  extinction using  the
library  written by  Simon~Krughoff\footnote{The library  is available
from  the  author upon  request.},  which  adopts  the SFD  dust  maps
\citep{1998ApJ...500..525S}    and    the    extinction    curve    by
\citet{1994ApJ...422..158O}.    Following  the  SDSS   convention  the
spectra   are  expressed  in   vacuum  wavelengths.    Flux  densities
$f_{\lambda}$               are              expressed              in
$10^{-17}$~ergs~s$^{-1}$~cm$^{-2}$~\AA$^{-1}$.

\section{Variability Amplitude: \fvarm} \label{section:fvar}

We adopt the dimensionless  variability measure, \fvarm, for $N$-epoch
repeated                                                   observations
\citep[e.g.,][]{1997ApJS..110....9R,2001sac..conf....3P}.    It  is  a
fractional root-mean-squared variability amplitude defined as

\begin{equation}
\fvar = \frac{\sqrt{\sigma^2 - \delta^2}}{\langle f\rangle} \ ,
\end{equation}

\noindent
where $\sigma^2$ is  the variance of the flux,  $\delta^2$ is the mean
square uncertainty  of the  flux, and $\langle  f\rangle$ is  the mean
flux

\begin{eqnarray}
\sigma^2  &   =  &   \frac{1}{N}  \sum_{i=1}^{N}  {(f_{i}   -  \langle
      f\rangle)}^{2} \ ,   \\  \delta^2  &   =  &   \frac{1}{N}
     \sum_{i=1}^{N} \delta_{i}^2 \ ,  \\ \langle  f\rangle &  = &
		 \frac{1}{N} \sum_{i=1}^{N} f_{i} \ .
\end{eqnarray}

\noindent
The numerator, proposed by \citet{1979A&AS...35..391B}, was termed the
square-root of the ``excess variance'' \citep{2003MNRAS.345.1271V}, in
the sense that  the flux uncertainty is subtracted  in quadrature from
the variability.   In the  calculation of the  continuum \fvarm  \ the
wavelength  intervals  encompassing  emission-line center  wavelengths
$\pm$~280~\kms \  are excluded (using the vacuum  wavelength values in
the  line list\footnote{\citet{2002AJ....123..485S}.}  adopted  by the
SDSS, and excluding the lines Ca~K and H, H$\delta$, the G band around
4306~\AA,   Mg$\lambda$5177,  Na$\lambda$5896,  and   the  \ion{Ca}{2}
triplet  that  usually  appear   as  absorptions).   The  mean  square
uncertainty for the spectrum at each epoch ($\delta_{i}^2$ for a given
$i$)  is  calculated by  quadrature  summation  of  the pipeline  flux
uncertainty  per pixel,  over all  valid  pixels.  It  is therefore  a
variance in flux based on  photon statistics. A valid pixel is defined
to be  a pixel that is  not flagged as  bad.  The flags are  listed in
\S\ref{section:bandvsoiii}.   \fvarm \ is  calculated in  the observed
frame with  wavelengths \obswaverange  \ for each  object and  for all
spectral types.

When  calculating \fvarm \  for narrow-line  AGNs, since  any possible
host-galaxy contribution is not explicitly subtracted from an observed
spectrum,  the  full observed  spectrum  is  used  in calculating  the
denominator  $\langle  f\rangle$.   As  the  numerator  term  ``excess
variance'' does not contain the non-variable host-galaxy contribution,
the  resultant  \fvarm  \  is  therefore  a lower  limit  on  the  AGN
variability.

If  the excess  variance  $\sigma^2 -  \delta^2  < 0$,  we assume  the
variability is zero for the object,  but continue to include it in the
analysis. In this  work, $N = 2$.  We note that  any non-zero \fvarm \
is  referred to as  a ``variability''  measurement, regardless  of the
origin being physically interesting or otherwise.

\section{Refining Spectroscopic Calibration} \label{section:recalibration}

The  principal steps  in  the spectrophotometric  calibration of  each
spectrum in the SDSS are: a wavelength-dependent calibration using the
observed  standard stars  on  the same  plate;  and the  tying of  the
absolute flux  scale to the observed  PSF magnitudes of  stars, on the
same   plate  \citep{2008ApJS..175..297A}.    The   resultant  average
uncertainty     in     the     spectrophotometry    for     DR6     is
\sdssspectrophotouncertainty  \  \citep{2008ApJS..175..297A}.  As  any
variability in  the sources from our  sample is expected to  be of low
amplitude, the  calibration of the  pipeline spectra must  be improved
using the approaches described below.

\subsection{Wavelength-dependent correction}  

\begin{figure}\begin{center}
\includegraphics[width=5in,angle=0]{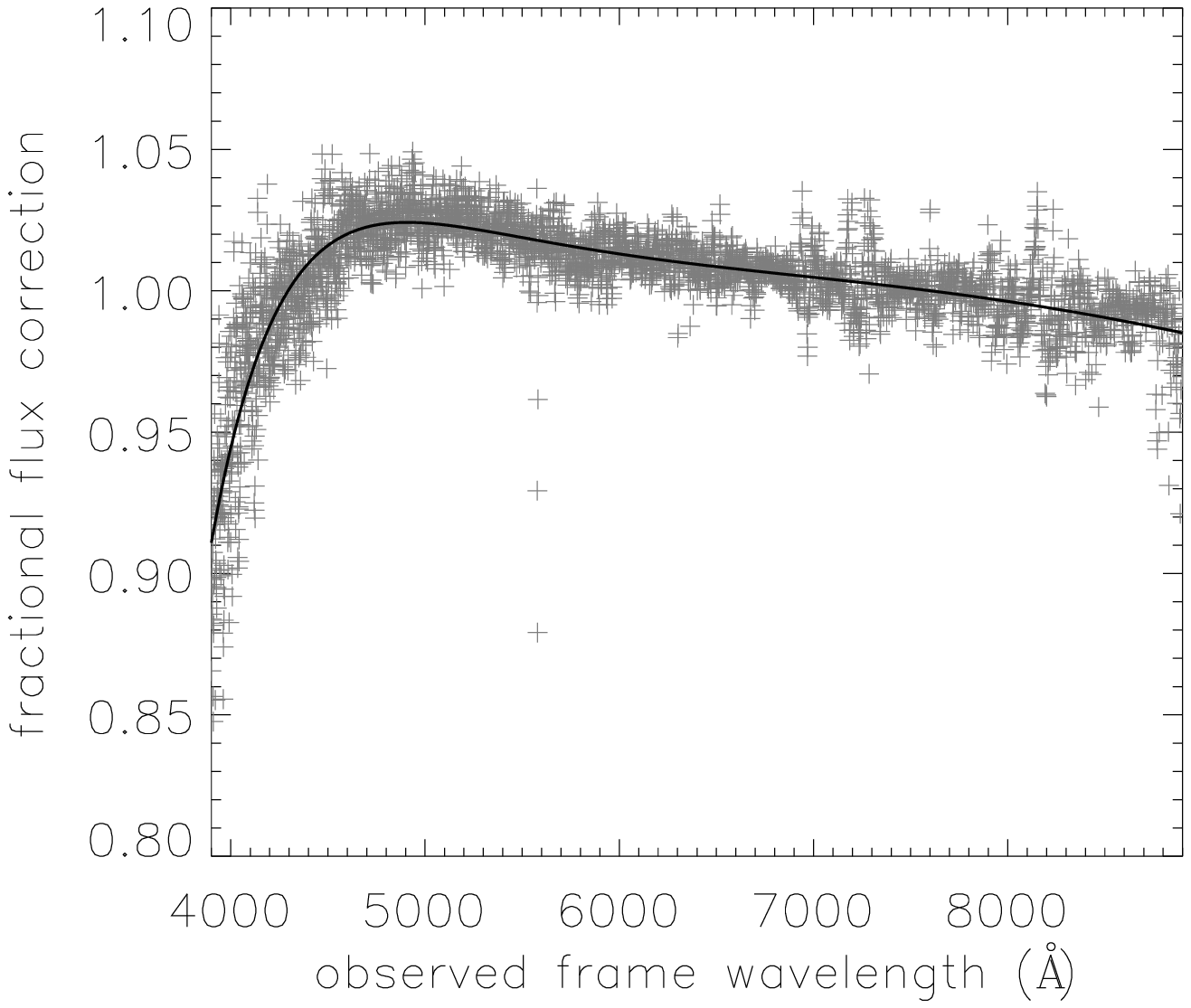}
\caption{The calibration spectrum (gray crosses) of the spectroscopic
plate  390  and  its  5th-order  polynomial  fit  (black  line).   The
correction in  flux density  between two epochs  is a function  of the
observed-frame wavelength.}
\label{fig:corrslope_plate390}
\end{center}\end{figure}

We adopt the method developed by \citet{2005ApJ...633..638W} to remove
systematics which may be present  in each pair of galaxy spectra taken
on  separate MJDs.   Flux-density  corrections as  a  function of  the
observed-frame  wavelength  in  the   range  of  \obswaverange  \  are
calculated  by a  linear fit  (tied to  the origin)  between  the flux
density  ($f_{\lambda}$) of  all  the objects  at  epoch~2 (the  later
epoch)  and   those  at  epoch~1.   This  calculation   results  in  a
calibration  spectrum for one  plate.  We  adopt a  slightly different
methodology from \citet{2005ApJ...633..638W}  in which stellar spectra
in  each plate  were  used.   Our calibration  spectrum  per plate  is
calculated using  all galaxies  present in that  plate, as we  want to
maximize the number of objects in generating the calibration spectrum,
and  avoid potential  systematic  effects when  applying point  source
calibration to  extended sources.   The underlying assumption  is that
the  majority  of   the  galaxies  observed  in  a   given  plate  are
non-variable in time across the two epochs.  Even if the galaxies were
in fact variable, the method  is still applicable because the variable
amplitude of several hundred galaxies  is not expected to be in phase.
An     example     calibration      spectrum     is     plotted     in
Fig.~\ref{fig:corrslope_plate390}.  Typically  the calibration spectra
are smooth.  To focus on  low-order corrections to the galaxy spectra,
and  to   remove  noise  in   the  calibration  spectrum,   we  follow
\citet{2005ApJ...633..638W} and set the final calibration spectrum per
plate  as  its  5th-order  polynomial least-squares  fit.   By  visual
inspection this  functional form  works well in  removing higher-order
spurious spectral features.

When constructing the calibration spectrum  we drop any plate in which
 there are less than \npairminCali  \ pairs of galaxy spectra. We then
 refine each and every spectrum present  in a given plate by using the
 calibration spectrum, if available.

\subsection{Wavelength-independent correction} \label{section:bandvsoiii}

After  removing wavelength-dependent  systematics, each  spectral pair
can,   in  principle,   be  subjected   to   a  wavelength-independent
normalization in their flux densities.  In order to obtain an accurate
variability  measurement of  each spectral  pair, the  total continuum
fluxes of the two spectra at a chosen common wavelength region are set
equal.  Specifically, for  each pair the galaxy spectrum  at the epoch
of lower S/N is  shifted in a wavelength-independent fashion according
to

\begin{equation}
{f_{\lambda}}^{\mathrm{(low   \,   \rm S/N)}} {\mathrm{(corrected)}}
=   C_{\mathrm{band}}   \, 
{f_{\lambda}}^{\mathrm{(low \, \rm S/N)}} {\mathrm{(original)}} \ ,
\end{equation}

\noindent
in which  $C_{\mathrm{band}}$  is  a  constant for  each  spectral  pair,
calculated as

\begin{equation}
C_{\mathrm{band}} = {F_{\lambda \in {\mathrm{band}}}}^{\mathrm{(high \,
\rm S/N)}}  / {F_{\lambda \in  {\mathrm{band}}}}^{\mathrm{(low \, \rm
S/N)}} \ ,
\end{equation}

\noindent
where  $F_{\lambda  \in  \mathrm{band}}$  is  the total  flux  in  the
restframe wavelength range (which  is referred to as ``rest-band'' for
convenience).    In  this  work   the  rest-band   is  chosen   to  be
6450~\AA~$\pm~520$~\kms,  as  such  no  prominent emission  lines  are
present  in the  galaxy  spectra\footnote{When  refining QSO  spectra
which are located at a wide range of redshift, an {\it observed frame}
wavelength range 6450~\AA~$\pm~520$~\kms \ is used.}.  Only the galaxy
spectra in the  first epoch are re-calibrated because  it is typically
the existence  of low S/N data  that results in  a second observation.
The  combined  procedure of  systematics  removal  and restframe  band
scaling is referred to ``SYS+BAND'' in the following discussion.

Given an initial sample (\S\ref{section:data}, \npairtotal \ objects),
we   impose  S/N   cuts   using  the   rest-band   measures  (or   the
$[$\ion{O}{3}$]\lambda$5008  (vacuum) line,  described  later when  we
discuss  absolute  flux calibration).   In  this  case  we reject  any
spectral pair  if there are  one or more  pixels flagged in  either or
both spectra in the wavelength  region of interest, and we require the
flux  to  be  larger  than  zero  at both  epochs.   The  flags  under
consideration  are  NOPLUG,  BADTRACE,  BADFLAT,  BADARC,  MANYBADCOL,
MANYREJECT,  LARGESHIFT, FULLREJECT,  SCATTERLIGHT,  CROSSTALK, NOSKY,
BRIGHTSKY,   NODATA,    COMBINEREJ,   BADFLUXFACTOR,   BADSKYCHI   and
REDMONSTER.   All of  the spectral  flux or  S/N calculations  in this
paper  are carried out  with pixel-masking  using these  flags.  After
this  S/N criteria,  the number  of objects  in the  galaxy  sample is
reduced to \npair \ (Table~\ref{tab:plate}).

\begin{figure}\begin{center}
\includegraphics[width=4.5in,angle=0]{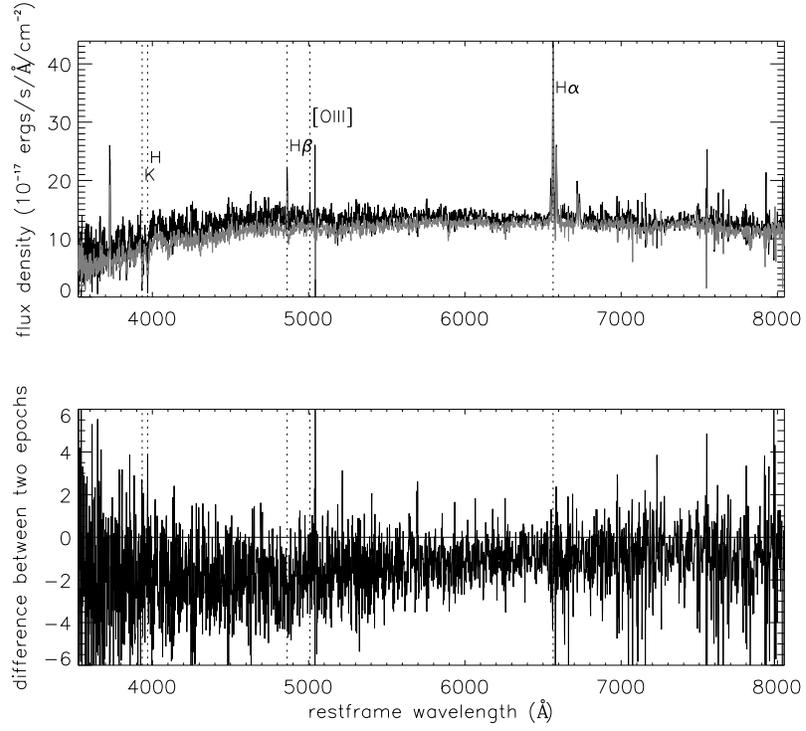}
\caption{An  example  star-forming   galaxy  without  any  calibration
refinement ($\fvar =  0.058$).  Top: spectra at the  first (black) and
the second (gray) epoch.  Bottom: difference spectrum.}
\label{spec_284_80160781273923584_before}
\end{center}\end{figure}

\begin{figure}\begin{center}
\includegraphics[width=4.5in,angle=0]{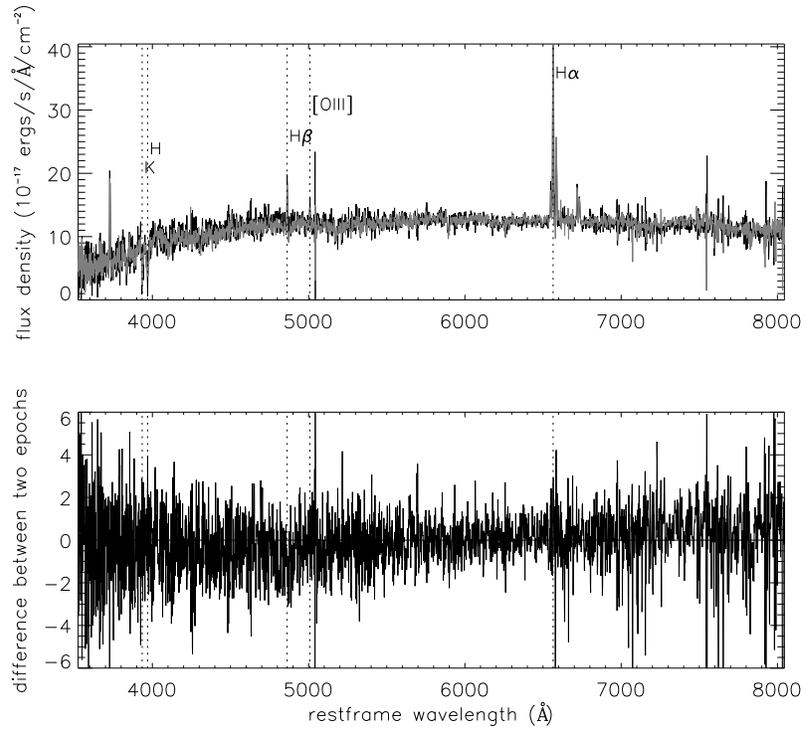}
\caption{The           same           galaxy           as           in
Fig.~\ref{spec_284_80160781273923584_before},   which  shows  zero
continuum  variability ($\fvar  = 0$)  after the  SYS+BAND calibration
refinement.  Top: spectra  at the first (black) and  the second (gray)
epoch.  Bottom: difference spectrum.}
\label{fig:spec_284_80160781273923584_after}
\end{center}\end{figure}

In           Fig.~\ref{spec_284_80160781273923584_before}          and
\ref{fig:spec_284_80160781273923584_after}   we  show   the  restframe
galaxy  spectra in two  epochs from  the SDSS  pipeline and  after the
calibration  refinement,  respectively.   Firstly,  we note  that  the
pipeline performs  well, that  the fractional difference  between both
epoch is  about 10\% at  the shortest wavelength, which  is sufficient
for  many spectral  analyses involving  emission or  absorption lines.
However,               continuum              residual              in
Fig.~\ref{spec_284_80160781273923584_before} is present, which induces
a variability ($\fvar =  0.058$).  After the calibration refinement we
are able to obtain a zero continuum variability.

For comparing with the rest-band scaling previously discussed, we also
adopt     the    absolute     flux     calibration    described     in
\citet{1998ApJ...501...82P}.  In  this procedure, the  spectrum in the
low S/N  epoch in the  duo is scaled wavelength-independently  so that
its total flux in  $[$\ion{O}{3}$]\lambda$5008 (vacuum) is the same as
that  of   the  spectrum  in   the  high  S/N  epoch.    The  physical
justification is  that the $[$\ion{O}{3}$]\lambda$5008  line generally
appears as a  narrow line arising from a  low-density region, which is
believed to  be non-variable in  time \citep[][]{1995PASP..107..579P}.
We note that the average  $[$\ion{O}{3}$]$ EW of galaxy spectra in the
full DR6 catalog is \oiiiavgsdssdrsix,  so for some objects mainly the
continuum  flux is  measured and  is expected  to be  non-variable as
well.   The combined  procedure  of systematics  removal and  absolute
$[$\ion{O}{3}$]$ calibration is  called ``SYS+OIII'' in the following.
The region of influence is set to be 5008~\AA~$\pm~280$~\kms.

When  judging  which  method  performs better  in  analyzing  spectral
variability,  we  consider the  better  method  to  be the  one  which
minimizes the mean, one standard deviation of the mean (\sdom) and the
one standard deviation sample scatter ($\sigma$) of the variability at
a fixed spectral S/N. These  statistics for the galaxy sample by using
the  SYS+BAND, SYS+OIII  and other  refinement methods  are  listed in
Table~\ref{tab:fvar}.  At  S/N \SNmin  \ the SYS+BAND  method produces
smaller mean and scatter in  \fvarm.  The difference is not related to
the difference in the width of the rest-band being used (520~\kms \ vs
280~\kms),  as   the  SYS+BAND  method  is   also  re-performed  using
6450~\AA~$\pm~280$~\kms,  the same  width as  in the  SYS+OIII method.
These statistics remain lower in value than those in the SYS+OIII even
for   this   reduced  interval.    The   result   suggests  that   the
$[$\ion{O}{3}$]\lambda$5008 line  in some  spectra may be  impacted by
the effect of seeing so that the total observed line flux is different
for the two  epochs.  The other possible reasons  for this discrepancy
are: the total  line flux is sampled at only a  few wavelength bins in
the medium resolution SDSS spectra,  or the line may exhibit intrinsic
variability.

\begin{figure}\begin{center}
\includegraphics[width=7in,angle=0]{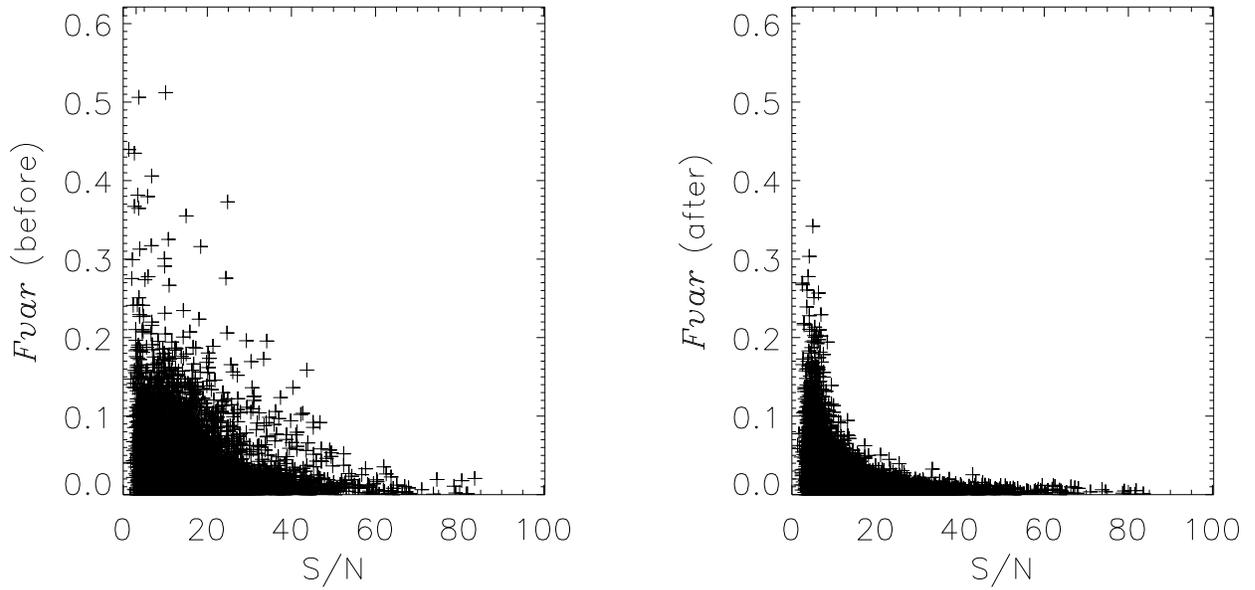}
\caption{\fvarm \ vs.   S/N of galaxy spectra before  (left) and after
  (right) the SYS+BAND calibration  refinement.  The number of objects
  in each spectroscopic plate is listed in Table~\ref{tab:plate}.}
\label{fig:fvar_sn_before_after}
\end{center}\end{figure}

To further test the method  SYS+BAND, we divide the galaxy sample into
RED and BLUE galaxies by using the eClass classification (described in
\S\ref{section:eclass}).  The RED spectra  are defined as objects with
classification  parameter   $\phi_{KL}  \geq  0^{o}$,   and  the  BLUE
galaxies,  $\phi_{KL}  <  0^{o}$.   For each  plate,  the  calibration
spectrum is calculated  using only the RED or  BLUE galaxies, which in
turn is  used to  re-calibrate galaxies of  all spectral types  on the
same  plate.  The  purpose behind  this  is to  determine whether  the
refinement  method  SYS+BAND  would  be  biased  by  the  presence  of
potentially  different galaxy spectral  types in  the plate.   We find
that using  a specific  spectral type does  not cause  the variability
statistics  to  change  (Table~\ref{tab:fvar}).   We  therefore  adopt
SYS+BAND  as the  calibration  method, and  use  all available  galaxy
spectral  types when calculating  the calibration  spectrum throughout
this  work.  The ``before  and after''  of this  approach is  shown in
Fig.~\ref{fig:fvar_sn_before_after}, where we  see that the scatter in
\fvarm  \  vs.   S/N  of  our  galaxy  sample  is  reduced  after  the
refinement.  Comparing  the average  variability of our  galaxy sample
using the SYS+BAND approach with  that from the original SDSS spectral
reductions (Table~\ref{tab:fvar})  we find that  the spectrophotometry
is improved by a factor of \specimprovement.

\subsection{Plate-to-Plate Difference}

\begin{figure}\begin{center}
\includegraphics[width=4in,angle=0]{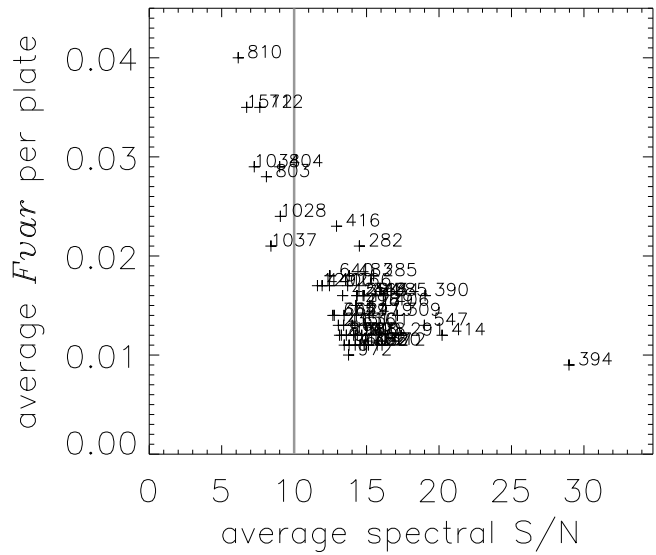}   
\caption{The  average  variability  of  galaxy spectra  over  a  given
spectroscopic  plate plotted  versus  the average  spectral S/N.   The
spectra at S/N  \SNmin \ are selected in  the subsequent definition of
variable and non-variable candidates to minimize variation in \fvarm \
among plates.}
\label{fig:fvar_sn_plate}
\end{center}\end{figure}

There are plate-to-plate differences  seen in the above definitions of
variable and non-variable sources,  that is, the best-fit coefficients
in   Eqn.~\ref{eqn:expenv}  are  in   general  plate   dependent.   In
Fig.~\ref{fig:fvar_sn_plate}  we show $\langle  \fvar \rangle$  of the
galaxy  spectra vs.   average spectral  S/N, per  plate.   The average
\fvarm \ per plate is larger at lower overall spectral S/N, suggesting
in those plates the variability  measurement is likely affected by the
presence  of noise in  the spectra.   By dropping  individual spectral
pairs with  spectral S/N $< 10$  at the second epoch,  the average and
the sample scatter in \fvarm \ are reduced to \fvarMeanDropLowSN \ and
\fvarSDDropLowSN, respectively.

\section{Galaxy Variability} \label{section:vari}

\subsection{Defining Variables}

Following \cite{2005ApJ...633..638W} we define the variable candidates
by objects having

\begin{equation}
F_{\mathrm{var}}(\rm     S/N)    >    b_0     \exp    \left({\frac{\rm
S/N}{b_1}}\right) + \, b_2 \ . \label{eqn:expenv} 
\end{equation}

\noindent
The S/N  of each  spectrum is  calculated in the  second epoch  and by
using    all     of    the     valid    pixels    as     defined    in
\S\ref{section:bandvsoiii}.   The   constants  $b_0,  b_1,   b_2$  are
determined by least-squares fitting  the above exponential form to the
binned      data     ($\langle      \mathrm{S/N}\rangle$,     $\langle
F_{\mathrm{var}}\rangle    +    n\,\sigma(F_{\mathrm{var}})$),   where
$\sigma(F_{\mathrm{var}})$ is  one standard  deviation of \fvarm  \ in
each bin, and  for $n = 0, 1,  2, 3, 4$. Typically 10 bins  in S/N are
used.

The variable  candidates (for galaxies,  stars and QSOs) in  this work
are  defined  as objects  with  \varinSigma  \  detection. We  do  not
sub-divide the distribution for detections greater than \variMaxSigma,
though in the future we plan to explore in more detail the tail of the
variable distributions.  For comparison we have also defined variables
as objects located in a fixed upper percentile of \fvarm \ in the plot
\fvarm \ vs.   S/N.  The \varinSigma \ cut is  equivalent to a $2-6$\%
upper percentile for a given S/N (Table~\ref{tab:percentile}).

\subsection{Continuum} \label{section:fvarvssn}

To explore any dependence  of spectral variability or repeatability on
the spectral type  of galaxies, we divide the  galaxy sample using two
different      spectral       classification      schemes:      eClass
\citep{1995AJ....110.1071C} and the ``BPT'' diagrams or the Osterbrock
diagrams  \citep{1981PASP...93....5B,1987ApJS...63..295V}.   These two
schemes are motivated by different spectral features.  The eClass is a
spectral  type describing  the steepness  of the  continuum  slope and
other higher-order spectral features  by using a linear combination of
several eigenspectra for each galaxy spectrum.  The eClass parameters,
$\phi_{KL}$ and $\theta_{KL}$, are  related to the coefficients in the
eigenspectra   expansion   (called   eigencoefficients),   which   are
calculated by  the Spectro1d pipeline in the  SDSS.  Specifically, the
angle   $\phi_{KL}$  correlates   with  the   H$\alpha$   emission  EW
\citep{2003MNRAS.343..871M}  which is an  indicator of  star formation
rate.  The angle $\theta_{KL}$ discriminates galaxy spectra exhibiting
post-starburst                                               activities
\citep{1995AJ....110.1071C,2004AJ....128..585Y}.

On the other hand, the BPT diagrams separate the narrow-line AGNs from
galaxies  such as the  star-forming galaxies  by emission  line ratios
which indicate the physical conditions of gas inside the galaxies such
as   the  electron   density  and   the  mean   level   of  ionization
\citep{1989agna.book.....O}.   The  objects   in  these  two  spectral
classification  schemes  are  not  mutually exclusive.   We  hereafter
present results on variability within both kinds of classification.

\subsubsection{eClass} \label{section:eclass}

The spectral variability  as a function of S/N  per mean eClass galaxy
type        \citep{1995AJ....110.1071C}       is        shown       in
Table~\ref{tab:fvarFullType},   and  Table~\ref{tab:fvarVariType}  for
only  the variable  candidates.  The  mean eClass  spectra  are galaxy
spectra where each is averaged  over a non-trivial size of subspace of
the    two   eClass    parameters   $\phi_{KL}$    and   $\theta_{KL}$
\citep{2004AJ....128..585Y}.  Limited by the number of spectral pairs,
\npairECLASSEF \  in types E or  F, only the types  A, B, C  and D are
considered in  the variable candidates (Table~\ref{tab:fvarVariType}).
The continuum of the mean galaxy spectrum is reddest in the type A and
bluest in the type D.

Firstly, we note that the average variability amplitude are consistent
among different eClass types to within 1--2$\sigma$ sample scatter. If
we assume  a priori that some  populations of galaxies do  not vary in
time, then a spectral repeatability  can be assigned to the population
with the smallest mean and scatter in \fvarm, that is, the mean eClass
type~B.   The spectral  features  of  this class  are  similar to  the
elliptical   galaxies   in   the   atlas   of   nearby   galaxies   by
\citet{1992ApJS...79..255K}.   The  repeatability,  taken  to  be  the
average   \fvarm,  is  \repeatabilitymin   \  in   the  observed-frame
wavelength range \obswaverange.

\subsubsection{Emission-line galaxies: narrow-line AGN
  vs. star-forming} \label{section:osterbrock}

Emission-line galaxies are considered  by using the BPT diagram, using
the       starburst      theoretical       modeling       line      by
\citet{2006MNRAS.372..961K}.   We  consider   the  types:  Seyfert  2,
star-forming, composite and  LINERs \citep{1980A&A....87..152H}. We do
not  find  any   LINER  in  our  galaxy  spectral   pairs,  using  the
classification           criteria           by          \citet[][their
Eqn.~15]{2006MNRAS.372..961K}              which              involved
$[$\ion{O}{1}$]$$\lambda{6302}$.   We   hereby  consider  the  diagram
log$_{10}$$([$\ion{O}{3}$]$/H{$\beta$})                             vs.
log$_{10}$($[$\ion{N}{2}$]$/H{$\alpha$})                  \citep[Fig.~5
of][]{1981PASP...93....5B}.

In the  first step of  this analysis the  EWs are taken from  the SDSS
reduction.   Galaxies with  any  one  of the  four  lines (\Hbeta\  at
$4863$~\AA,   [\ion{O}{3}]$\lambda$5008,   \Halpha\   at   $6565$~\AA,
[\ion{N}{2}]$\lambda$6585) having  restframe EW smaller  than \ewMin \
are rejected (in the SDSS convention EW $> 0$ for emission lines);
and at least a \nsigmaMin \  detection is required in each of the four
lines.  Restframe stellar absorption  of the related hydrogen emission
lines in the  BPT diagram is corrected for by  a constant increment of
\stellarabs  \  \citep{2003ApJ...599..971H,2003ApJ...597..142M}.   Any
possible aperture  effect on the  emission lines is neglected  in this
paper, because the primary interest  of this work is the comparison of
line  strength  between two  epochs,  which  is  done using  the  same
aperture for each object in each epoch and will only be subject to the
difference in  seeing.  To correct on an  object-by-object basis would
require an  assumption of  an average line  strength as a  function of
radius  for each galaxy.   Although the  true line-flux  correction of
individual objects  should depend on  the distribution of the  gas and
its projected angle on the sky relative to the observational aperture,
under the  typical sky conditions  in the SDSS  observation \citep[the
median  PSF   width  =   1.4\arcsec  \,  in   the  $r$   band,  Fig.~4
of][]{2003AJ....126.2081A} and  given the  diameter of the  fiber, the
seeing-induced aperture correction factor  to [\ion{O}{3}] is close to
unity, based on a nearby ($z  = 0.009$) Seyfert 1 with a well-resolved
($\sim{10}$\arcsec)       narrow-line       region       \citep[Fig.~3
of][]{1992A&A...266...72W}.     This    means   that    seeing-induced
differences in the [\ion{O}{3}] flux  between any two epochs should be
small. Further, the SDSS  3\arcsec \, diameter fiber spectroscopy does
not  cause substantial  aperture  bias in  classifying galaxy  spectra
based on emission lines \citep{2003ApJ...597..142M}.

In Table~\ref{tab:fvarFullType}  we compare the  variability amplitude
between the star-forming galaxies and the narrow-line AGNs in the full
galaxy  sample,  and  Table~\ref{tab:fvarVariType}  for  the  variable
candidates.  It  is a little  surprising that the  average variability
amplitude at S/N \SNmin \  is \fvarfullsampleSF \ for the star-forming
galaxies,  slightly   higher  than  that  of   the  narrow-line  AGNs,
\fvarfullsampleAGN.  The difference is bigger than 1\sdom \ (i.e., the
measurement of the \fvarm \  difference is reliable), but smaller than
the \onesigma \ sample scatter,  indicating the \fvarm \ of both types
are consistent with each other.

In Table~\ref{tab:contvari} we show the average absolute AB magnitudes
in $g,r$  and $i$ bands of  variable candidates in the  dim phase, and
the sample-averaged  magnitude difference  between the dim  and bright
phases,  per spectral type.   The magnitudes  are calculated  for each
spectrum, by convolving each  restframe-shifted spectrum with the SDSS
filter  response curves  for extended  sources  at an  airmass of  1.3
\citep{1996AJ....111.1748F},   and    in   the   AB    filter   system
\citep{1983ApJ...266..713O}.  The  average difference in  the $g$-band
magnitudes ranges from \contvarirangegdiff \ depending on the spectral
type, but they are consistent among different spectral types to within
\onesigma   \   sample    scatter   of   the   magnitude   difference,
$\sigma(\Delta{M})$, a  measure of the  sum of the sample  scatters in
both the distributions  of $M(\textrm{dim})$ and $M(\textrm{bright})$.
For the  narrow-line AGNs, no  obvious continuum variability  is found
(e.g., \dgSEYFERT \ in the $g$ band).

We  note that  the distribution  of  \fvarm \  in the  full sample  is
non-Gaussian. Using the \varinSigma \ cut for the variable candidates,
for a  Gaussian distribution  one would expect  the ratio  between the
number of variable candidates and that of the whole sample to be

\begin{equation}
\frac{1 - \mathit{erf}(3/\sqrt{2})}{1 - \mathit{erf}(-\infty)} \ ,
\end{equation}

\noindent
or $0.13$~\%,where  $\mathit{erf}(z)$ is the error function,  and $1 -
\mathit{erf}(z)$  is  the complementary  error  function  of the  form
$(2/\sqrt{\pi})\int_{z}^{\infty}    \,   e^{-   t^{2}}    dt$.    From
Table~\ref{tab:percentile}, we  see that typically at  various S/N the
\fvarm  \ of  the  variable candidates  span  a longer  tail ($1.53  -
3.44$~\%).

\subsubsection{Structure Function of the Continuum}

The  characteristic time scale  of variability  is usually  studied by
constructing     the    structure     function     (\structf,    e.g.,
\cite{1985ApJ...296...46S}, \cite{1992ApJ...396..469H}; and references
therein), the  variability amplitude  \fvarm \ as  a function  of time
lag. Although no obvious continuum  variability is found in our sample
of galaxies on average, the  \structf \ is nonetheless constructed for
the galaxy sample.   Through the \structf \ analysis,  an absence of a
time scale would be a cross-check  of the lack of variability, and the
comparison   between   the  \structf   \   of   stars   and  of   QSOs
(\S\ref{section:comparestarqso})  with  that of  the  galaxies can  be
carried out.

\begin{figure}\begin{center}
\includegraphics[width=5in,angle=0]{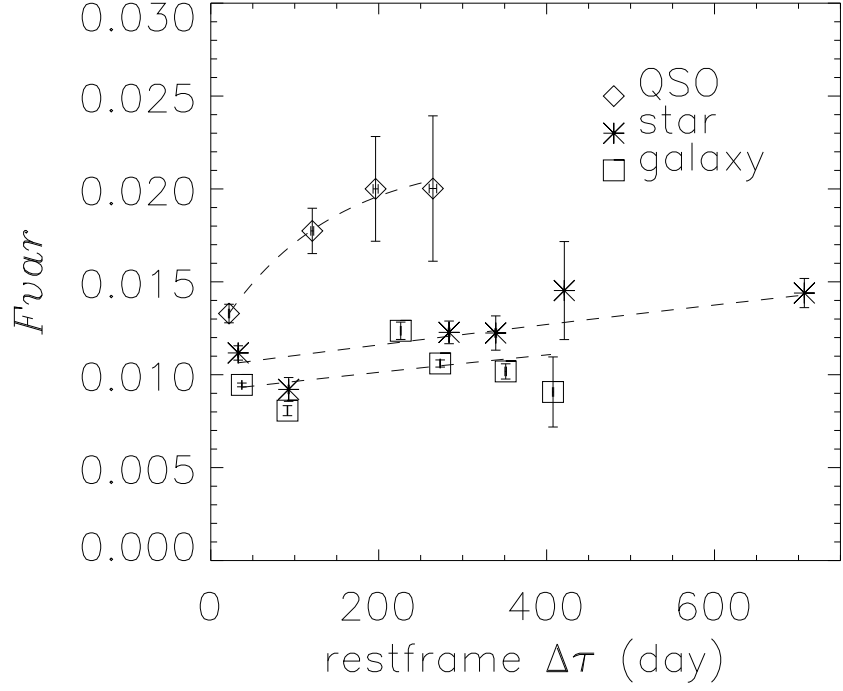}
\caption{Restframe structure function of the QSOs, stars and galaxies.
  Only spectra with S/N \SNmin \ are shown.  Error bar is 1\sdom \ for
  each axis.  Dotted line shows  the best-fit structure function of an
  exponential functional form in \dtaurest.}
\label{fig:sf_80_10_galaxy_qso_star}
\end{center}\end{figure}

\begin{figure}\begin{center}
\includegraphics[width=5in,angle=0]{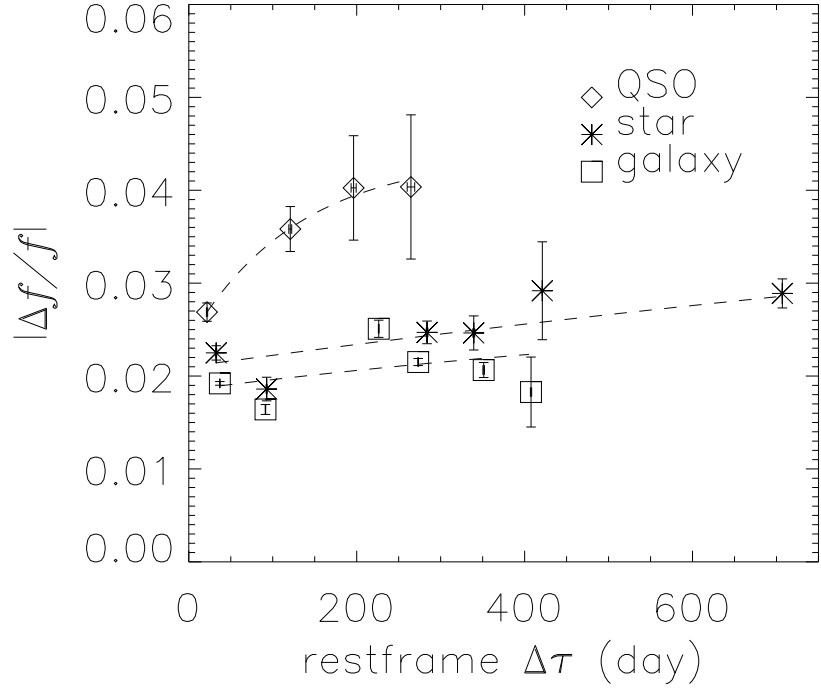}
\caption{Restframe       structure       function.       Same       as
Fig.~\ref{fig:sf_80_10_galaxy_qso_star}   except  that   the  fraction
change in total flux $|$\dfoverfm$|$ \ is used instead of \fvarm.}
\label{fig:sf_dfoverf_80_10_galaxy_qso_star}
\end{center}\end{figure}

The restframe  \structf, \fvarm \  vs.  \dtaurest, of all  galaxies in
the        galaxy        sample        is        illustrated        in
Fig.~\ref{fig:sf_80_10_galaxy_qso_star},   where  $\Delta\tau_{\mathrm
rest} = \Delta\tau /  (1 + z)$, and $z$ being the  redshift of a given
galaxy. The  time lag $\Delta\tau$  is the difference between  the two
MJDs  of   observation,  MJD(epoch  2)$-$MJD(epoch   1).   No  obvious
time-scale is seen in the  above restframe \structf{s} of galaxies for
the  \dtauRestRangeGALAXY  \  days  (restframe)  sampled.   A  similar
conclusion   is  obtained   when   the  fractional   change  in   flux
($|$\dfoverfm$|$)     is     used      instead     of     \fvarm     \
(Fig.~\ref{fig:sf_dfoverf_80_10_galaxy_qso_star}),  where  $\Delta{f}$
is the difference in the  flux between two epochs, and $f$ is the
average flux over the two epochs. 

An exponential function of the form
\begin{equation}
\fvar  =  a \,  (1  -  b\,  e^{-\Delta\tau_{\mathrm rest}  /  \tau_s})
\label{eqn:sf}
\end{equation}

\noindent
is  fitted  to   each  structure  function,  where  \taus   \  is  the
characteristic variability time scale. Typically, a structure function
where          $b         =          1$          is         considered
\citep[e.g.,][]{1979A&AS...35..391B,1994ApJ...433..494T}.   Firstly we
relax $b$  to accommodate the  non-zero variability at zero  time lag,
obvious  in   Fig.~\ref{fig:sf_80_10_galaxy_qso_star}.   This  can  be
attributed   to  the  repeatability   of  the   continuum  variability
measurement (\S\ref{section:eclass}) in  eClass Type~B, where the mean
\fvarm \ is \repeatabilitymin.  The  best-fit \taus \ for the galaxies
is   \tausGALAXY.    Uncertainty  of   this   time   scale  is   large
(\tausUncertaintyGALAXY, with a  reduced \chisq \ of \tausChisqGALAXY)
because  the duration  of the  observation  is much  shorter than  the
best-fit \taus.

\begin{figure}\begin{center}
\includegraphics[width=5in,angle=0]{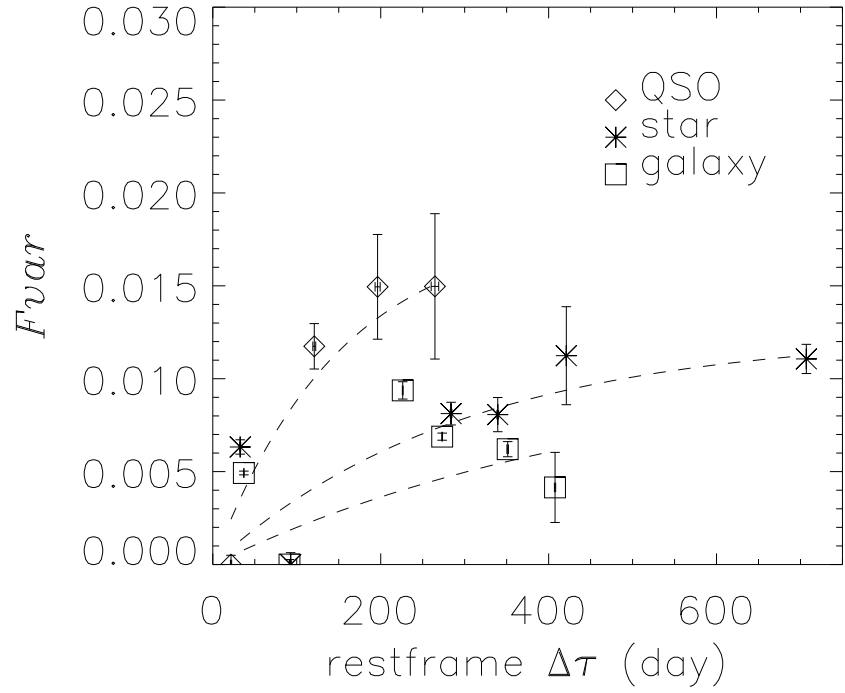}
\caption{Restframe structure function of the QSOs, stars and galaxies.
  Different  from Fig.~\ref{fig:sf_80_10_galaxy_qso_star}  which shows
  the total variability, here the  variability at the zero time lag is
  quadrature  subtracted, respectively  in each  spectral  type.  Only
  spectra with S/N \SNmin \ are shown.  Error bar is 1\sdom \ for each
  axis.   Dotted line  shows  the best-fit  structure  function of  an
  exponential functional form in \dtaurest.}
\label{fig:sf_80_10_galaxy_qso_star_quadsubtract}
\end{center}\end{figure}

We  then consider  a  typical $b  =  1$ structure  functional form  at
Eqn.~\ref{eqn:sf}.   By  assuming  \fvarm  \  at  the  zero  time  lag
(\fvarm$(0)$)  (the noise)  is uncorrelated  with the  intrinsic \fvarm
(the signal),  one can  obtain the intrinsic  \fvarm \  by subtracting
\fvarm$(0)$  from the  total \fvarm  \ in  a quadrature  fashion.  The
resultant       structure       function       is       shown       in
Fig.~\ref{fig:sf_80_10_galaxy_qso_star_quadsubtract}, and the best-fit
variability  time scales  in  Table~\ref{tab:taus_quadsubtract}, where
\fvarm$(0)$ is taken to be the  minimum of the total \fvarm. The large
best-fit  \chisq \  in stars  and galaxies  implies that  the best-fit
variability  time scales  are quite  uncertain. A  longer  duration of
observation would be ideal to improve the structure function. 

\subsection{Emission Lines} \label{section:linevari}

The BPT diagram is constructed  for the variable candidates to explore
the narrow emission-line  variability.  Again the emission-line ratios
$\log_{10}$($[$\ion{N}{2}$]$/H{$\alpha$})                           vs.
$\log_{10}$($[$\ion{O}{3}$]$/H{$\beta$}) are considered.

In  measuring  the  EW  of  an  emission  line,  we  fit  to  a  given
continuum-subtracted  spectrum   a  single  Gaussian   function.   The
continuum   is  estimated   by  non-negative   least   square  fitting
\citep{LawsonHanson74},  provided  by \citet{2006IAUSS...3E..76D}  who
used     the     \citet{2003MNRAS.344.1000B}    stellar     population
1~\AA-resolution models.   In this way, the stellar  absorption in the
Balmer lines is implicitly taken  into account in the best-fit stellar
continuum.  In measuring the uncertainty of an EW, the following steps
are performed.   Firstly, the uncertainty in  the luminosity densities
of  the stellar  continuum  of  the spectrum  is  calculated by  using
$\delta{f_{\lambda}^c} =  \delta{f_{\lambda}}$, where $\delta$ denotes
the uncertainty of the continuum flux density ${f_{\lambda}^c}$ and of
the observed spectrum ${f_{\lambda}}$.   The uncertainty in each EW is
calculated    to   be    $\sum_{\lambda_R}    {\delta{(f_{\lambda}   -
f_{\lambda}^c)}/{f_{\lambda}^c}}  \, d\lambda$,  where  $\lambda_R$ is
the region of influence of the  emission lines, taken to be around the
line center $\pm  280$~\kms.  Finally, the uncertainty in  each of the
line  ratios \NIIHalpha  \ and  \OIIIHbeta  \ is  propagated by  using
common error propagation formula.

\begin{figure}\begin{center}
\includegraphics[width=4.5in,angle=0]{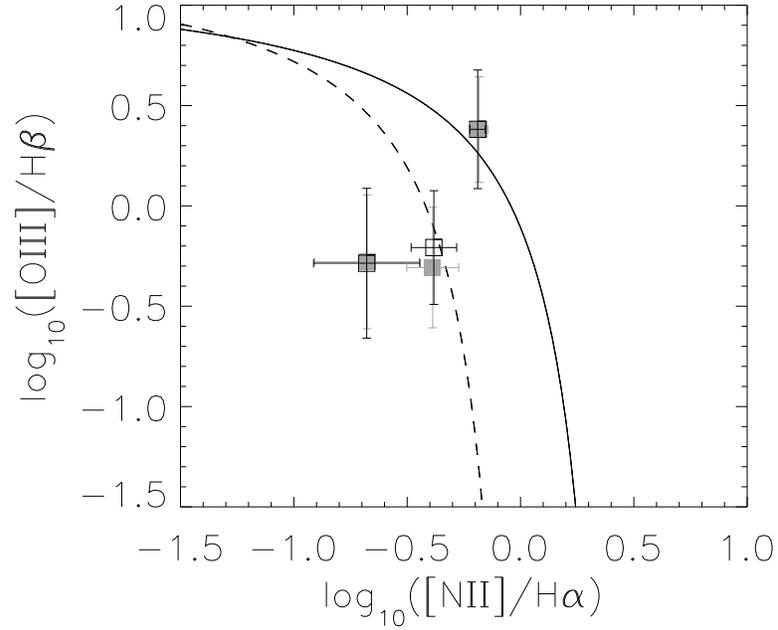}
\caption{The BPT  diagram of  the continuum variable  candidates.  The
object  type from  the lowest  to the  highest  [\ion{N}{2}]/\Halpha \
value  is: star-forming,  composite and  Seyfert 2.   In each  duo the
filled  square represents  the  average line  ratios  of the  brighter
objects,  and the  empty square  for the  dimmer ones.  The  solid and
dashed  lines  respectively  define  the Seyfert  2  and  star-forming
galaxies as  given by  \cite{2006MNRAS.372..961K}. The line  ratios in
all  object  types  do   not  show  substantial  change  between  both
epochs. The  error bar  represents the \onesigma  \ sample  scatter in
each phase. }
\label{fig:bpt_brg_dim}
\end{center}\end{figure}

\begin{figure}\begin{center}
\includegraphics[width=4.5in,angle=0]{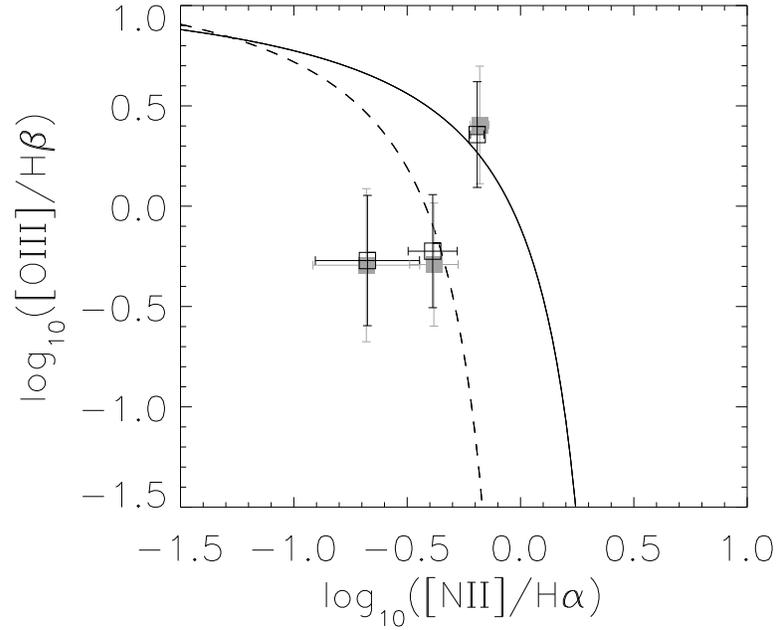}
\caption{Similar  to Fig.~\ref{fig:bpt_brg_dim}, but  in each  duo the
  filled square represents the average line ratios of the first epoch,
  and the empty square for  the latter epoch. The error bar represents
  the \onesigma \ sample scatter in each epoch.}
\label{fig:bpt_t1_t2}
\end{center}\end{figure}

\begin{figure}\begin{center}
\includegraphics[width=4.5in,angle=0]{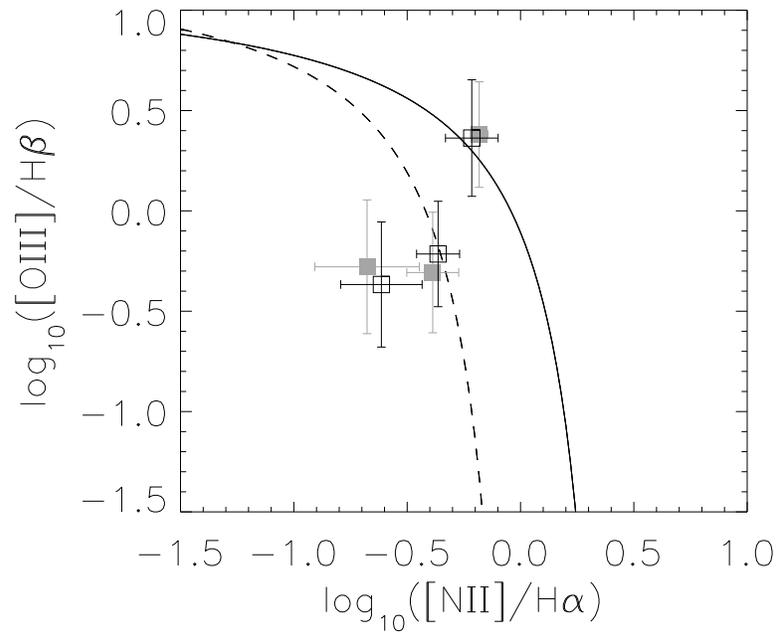}
\caption{Similar  to Fig.~\ref{fig:bpt_brg_dim}, but  in each  duo the
filled  square represents  the average  line ratios  of  the continuum
variable  candidates,  and  the  empty  square  for  the  non-variable
ones. Line ratios  in both cases are from the  bright phase. The error
bar represents the \onesigma \ sample scatter.}
\label{fig:bpt_brg_sigma_0_2_vs_sima_3_4}
\end{center}\end{figure}

The line ratios between the dim and bright phases of sources are found
to  be  located  very  closely   on  the  BPT  diagram,  as  shown  in
Fig.~\ref{fig:bpt_brg_dim}.    The  emission-line  ratios   and  their
uncertainties    for    various   object    types    are   given    in
Table~\ref{tab:linevari}.  The uncertainty  in the flux measurement is
small,  for example in  the star-forming  galaxies, which  indicates a
reliable  line  measurements.  For  both  line  ratios the  difference
between  the dim  and bright  phases  is typically  comparable to  the
\onesigma  \ sample scatter  in the  line-ratio difference  (i.e., the
last two columns of  Table.~\ref{tab:linevari}).  This is true for all
of the object types considered,  meaning we find no evidence of narrow
emission-line-ratio  variability  which is  above  \onesigma \  level.
When  dividing the  line-ratio measurements  by the  first  and second
epoch (Fig.~\ref{fig:bpt_t1_t2}) instead of the dim and bright phases,
we  also  see  no difference  between  the  two  that is  larger  than
\onesigma.  Further,  the line ratios  of the variable  candidates and
the                       non-variable                      candidates
(Fig.~\ref{fig:bpt_brg_sigma_0_2_vs_sima_3_4}) do not show substantial
difference  in the  BPT diagram,  true  for all  of the  emission-line
galaxy types.

The  above   analyses  are   also  repeated  with   the  intrinsically
de-reddened   spectra  using  the   best-fit  color   excess  (E(B-V))
(Yip~et~al.~2008 in  prep.), by the non-negative  least square fitting
of   stellar   population   models    on   the   SDSS   DR6   galaxies
\citep{2006IAUSS...3E..76D}.   Expectedly, each  of  the average  line
ratios remains unchanged to the \onesigma \ level (not shown), because
the central  wavelengths of the two  lines for a given  line ratio are
close and is thus insensitive to dust reddening. 

\section{Comparison with Star and QSO} \label{section:comparestarqso}

To evaluate further the  performance of the calibration refinement, we
refine the calibration  in stellar and QSO spectra  in a spectroscopic
plate using calibration spectrum constructed from galaxies of the same
plate. The procedure is described in \S\ref{section:recalibration}.

\subsection{Variability in SDSS Stellar and QSO spectra} 

Table~\ref{tab:fvar_star}  lists the  statistics  of \fvarm  \ in  our
sample  of   the  SDSS  stellar  spectra.   The   improvement  in  the
repeatability   after  the   SYS+BAND  refinement   is  a   factor  of
\specimprovementStar. The  average \fvarm  \ of stars  is found  to be
slightly higher than that of the galaxies.

Table~\ref{tab:fvar_qso}  lists  the statistics  of  \fvarm  \ in  our
sample of the SDSS QSO spectra. The QSOs are located within a redshift
range of  \qsoredshiftmin \  -- \qsoredshiftmax \  with an  average of
\qsoredshiftavg. Upon the \structf  \ analysis the characteristic time
scale of  the QSOs is  found to be \tausQSO  \ (Table~\ref{tab:taus}),
which agrees  with the  time scales commonly  found in  the literature
\citep[e.g.,][]{1996A&A...306..395C,2004IAUS..222..525I,2004ApJ...601..692V}
for the UV-optical variability.

The comparison between  the \structf \ of galaxies,  stars and QSOs in
Fig.~\ref{fig:sf_80_10_galaxy_qso_star},\ref{fig:sf_dfoverf_80_10_galaxy_qso_star}
shows that \fvarm(QSO) $>$ \fvarm(star) $>$ \fvarm(galaxy).

\subsection{Cross match with GCVS} 

The  full samples of  galaxies, stars  and QSOs  are matched  with the
General  Catalogue  of Variable  Stars  \citep[GCVS,][Vol.  I-III  and
references  therein]{1999yCat.2214....0K}  as  an  assessment  to  the
definition of variability.  The RA  and DEC are both matched to within
\radeccrossmatcharcsec, respectively.

Table~\ref{tab:gcvs} shows the objects that exist in both the GCVS and
our   samples.  As   expected,  all   of  the   matched   objects  are
spectroscopically classified as stars by  the SDSS.  Three of the five
GCVS  stars  are  defined  in  this work  to  be  variable  candidates
(\varinSigma  \ detection), hence  a completeness  of 60\%  when using
2-epoch observations to infer the stellar variability.

The above indicates that the variability of stars and galaxies in this
study could  in principle  be limited by  the small number  of epochs,
also pointed out by  \cite{2003MmSAI..74..978I}; and that the defining
of variable stars  by more than \threesigma \ over  the average in the
variability may need to be adjusted.   The fact that more than half of
the variable candidates are real  variable stars according to the GCVS
shows that the calibration refinement is successful in selecting those
objects.

\subsection{Cross match with SDSS Southern Stripe} 

Our full  samples of galaxies,  stars and QSOs are  also cross-matched
with the  variable catalog constructed  by \citet{2007AJ....134.2236S}
based on  the SDSS repeated imaging  (the Stripe 82).   The data cover
more observation epochs  (from 4 to 28, with an  average of 9), useful
for determining the completeness of our variable candidates based on 2
epochs. The  matched objects are found  to be either stars  or QSOs in
the  SDSS  spectral  classification, as  \citet{2007AJ....134.2236S}'s
catalog     is      for     un-resolved     objects      only.      In
Table~\ref{tab:varStripe82_star}  the matched  stars are  listed. Five
out of 22 sources matched to  the SDSS variable catalog are defined to
be variable stars  in this work (\varinSigma \  detection), implying a
23\% completeness when using 2-epoch observations to infer the stellar
variability.

For QSOs, because more  than \matchQSOSouthStripe \ matches are found,
only    the   number    of    found   objects    is   summarized    in
Table~\ref{tab:varStripe82_qso}, along with our detection significance
as  variable  candidates.   About  \matchQSOSouthStripeVari \  of  the
objects are considered  to be real variables in this  work. If some of
the narrow-line AGNs do vary in time, the completeness of the variable
candidates  from  our  current  2-epoch   sample  may  be  as  low  as
\matchQSOSouthStripeVari.

\section{Summary} \label{section:conclusion}

We probe the spectroscopic variability in the galaxies and narrow-line
AGNs in the optical  wavelengths using multi-epoch observations in the
SDSS.  In order  to detect the expected low  amplitude of variability,
we  compare  several   approaches  to  refine  the  spectrophotometric
calibration in the SDSS.  The  final calibration, in terms of the mean
and the standard deviation of the variability in our galaxy sample, is
a  factor  of  \specimprovement   \  better  than  the  official  SDSS
reductions.

Our sample of galaxy pairs  spans in the restframe $\sim{110}$~days on
 average, with  a maximum of  $\sim{700}$~days. The average  change in
 the narrow-line  AGN continuum flux,  when converted to  synthetic AB
 absolute magnitudes  using the SDSS  filters, is \dgSEYFERT \  in the
 synthetic  $g$ band,  where the  uncertainty is  propagated  from the
 uncertainty  in the  spectral  flux measurement.   The  fact that  no
 obvious  continuum  variability  is  found  is  consistent  with  the
 expectation       from        the       AGN       unified       model
 \citep{1993ARA&A..31..473A,1995PASP..107..803U}, in which some of the
 light-emitting (in our context, continuum-generating) structures in a
 narrow-line AGN, such as the  accretion disk, are obscured by a dusty
 torus.   Hence,   any  variability  in  such   structures  (cf.   the
 UV-optical variability  in QSOs in the  time scale of  1~year) is not
 observed.  This also provides  an empirical evidence for modeling the
 continua  of   narrow-line  AGNs  using   stellar  population  models
 \citep[e.g.,][]{2003MNRAS.346.1055K,2005AJ....129.1783H}.     Further,
 if  we use  this continuum  variability measurement  to set  an upper
 limit on the  AGN activities, then the ratio between  the flux of any
 varying spectral component, presumably related to AGN activities, and
 that of the host galaxy is at most \dgfluxSEYFERTUpperLimit.

By  comparing  the  narrow   emission-line  ratios  \NIIHalpha  \  and
\OIIIHbeta  \  between  the  defined  dim and  bright  phases  in  the
continuum  variable   candidates,  we  find  no   evidence  for  their
variability  to be substantially  larger than  the \onesigma  \ sample
scatter  in the line-ratio  difference.  This  is consistent  with the
common  view  in which  the  narrow-line  emissions  in the  AGNs  are
generated from  low-density clouds.  For example,  an electron density
$n_e  \sim{1,000}$~cm$^{-3}$ gives  rise to  a  hydrogen recombination
time  $\sim{100}$~years,  which  is  typically much  longer  than  the
duration of the observation.

We found  no evidence of  continuum variability in galaxies  of eClass
types  A, B, C,  D that  is larger  than the  variability uncertainty.
This agrees  with the expectation  that stellar light  dominates their
continuum, as in the narrow-line AGNs.

We tabulate the  upper limits of both the  continuum and emission-line
variability as  a function of spectral type,  using the classification
schemes  eClass and  the BPT  diagram.  These  values can  serve  as a
sanity check for researchers who study variability for other sources.

\subsection{Variability in Classification}

There were  few attempts in  examining the galaxy  classification from
the   point   of   view    of   variability.    A   recent   work   by
\cite{2001A&A...373...38B}  suggested   variability  is  an  efficient
method to  find narrow-line AGN.   However, this promising  result was
not subsequently confirmed by the authors \citep{2002A&A...390..439M}.
Our result on narrow  emission-line ratios being non-variable seems to
agree with the latter claim made by the authors.

\subsection{Next Steps}

So far  we have focused on  the average variability  properties of the
galaxies  and   narrow-line  AGNs.   We  are   investigating,  on  the
object-to-object basis, any dramatic variation in spectral features or
types.

\section{Acknowledgements}

We   thank  Brian~Wilhite,   David~Turnshek   and  Julian~Krolik   for
discussions.     We   thank   Brigitte~K\"{o}nig,    Amy~Kimball   and
Scot~Kleinman for discussions on  variable stars.  We thank Ani~Thakar
for discussions on  SDSS databases.  We thank the  referee for helpful
suggestions, and discussions on  narrow-line Seyfert 1 galaxies.  ASS,
RFGW, TB and CWY acknowledge support through grants from the W.M.~Keck
Foundation and the  Gordon and Betty Moore Foundation,  to establish a
program  of data-intensive  science at  the Johns  Hopkins University.
AJC acknowledges  partial support from  the NSF ITR award  0851007. LD
and  IC  acknowledge  support  from  grants  MSRC-2005-038,  NAP-2005/
KCKHA005 and MRTN-CT-2004-503929.

Funding  for the  SDSS and  SDSS-II has  been provided  by  the Alfred
P.  Sloan  Foundation, the  Participating  Institutions, the  National
Science  Foundation,  the  U.S.  Department of  Energy,  the  National
Aeronautics and Space Administration, the Japanese Monbukagakusho, the
Max  Planck Society,  and  the Higher  Education  Funding Council  for
England. The SDSS Web Site is http://www.sdss.org/.

The SDSS is  managed by the Astrophysical Research  Consortium for the
Participating  Institutions. The  Participating  Institutions are  the
American Museum  of Natural History,  Astrophysical Institute Potsdam,
University  of Basel,  University of  Cambridge, Case  Western Reserve
University,  University of Chicago,  Drexel University,  Fermilab, the
Institute  for Advanced  Study, the  Japan Participation  Group, Johns
Hopkins University, the Joint  Institute for Nuclear Astrophysics, the
Kavli Institute  for Particle  Astrophysics and Cosmology,  the Korean
Scientist Group, the Chinese  Academy of Sciences (LAMOST), Los Alamos
National  Laboratory, the  Max-Planck-Institute for  Astronomy (MPIA),
the  Max-Planck-Institute  for Astrophysics  (MPA),  New Mexico  State
University,   Ohio  State   University,   University  of   Pittsburgh,
University  of  Portsmouth, Princeton  University,  the United  States
Naval Observatory, and the University of Washington.

\clearpage

\clearpage

\clearpage

\clearpage

\clearpage

\clearpage

\renewcommand{\thefootnote}{\alph{footnote}}

\begin{center}
\scriptsize
\begin{longtable}{cccccc}
\caption{Spectroscopic plates used in the analysis of galaxies.} \\
\hline
  plate\tablenotemark{a,b} &  
  first epoch MJD  & 
  second epoch MJD  & 
  difference in MJD &
  \begin{sideways}  {total number of pair\tablenotemark{c}} \end{sideways} & 
  \begin{sideways} {number of pair\tablenotemark{d}} \end{sideways}
\\
\hline\endfirsthead
\endhead
\hline {Continued on next page}\\\hline\endfoot
\endlastfoot
266	&	51602	&	51630	&	28	&	447	&	356	\\
279	&	51608	&	51984	&	376	&	430	&	361	\\
282	&	51630	&	51658	&	28	&	346	&	278	\\
284	&	51662	&	51943	&	281	&	425	&	336	\\
285	&	51663	&	51930	&	267	&	417	&	317	\\
291	&	51660	&	51928	&	268	&	444	&	344	\\
296	&	51665	&	51984	&	319	&	448	&	364	\\
297	&	51663	&	51959	&	296	&	453	&	369	\\
300	&	51666	&	51943	&	277	&	386	&	288	\\
301	&	51641	&	51942	&	301	&	370	&	282	\\
304	&	51609	&	51957	&	348	&	398	&	317	\\
306	&	51637	&	51690	&	53	&	435	&	352	\\
309	&	51666	&	51994	&	328	&	371	&	287	\\
340	&	51691	&	51990	&	299	&	385	&	302	\\
348	&	51671	&	51696	&	25	&	313	&	240	\\
351	&	51695	&	51780	&	85	&	395	&	311	\\
385	&	51783	&	51877	&	94	&	326	&	259	\\
390	&	51816	&	51900	&	84	&	392	&	313	\\
394	&	51812	&	51913	&	101	&	347	&	267	\\
404	&	51812	&	51877	&	65	&	421	&	335	\\
406	&	51817	&	52238	&	421	&	319	&	239	\\
410	&	51816	&	51877	&	61	&	309	&	243	\\
411	&	51817	&	51873	&	56	&	398	&	318	\\
412	&	51931	&	52258	&	327	&	396	&	306	\\
413	&	51821	&	51929	&	108	&	351	&	274	\\
414	&	51869	&	51901	&	32	&	383	&	299	\\
415	&	51810	&	51879	&	69	&	366	&	277	\\
416	&	51811	&	51885	&	74	&	332	&	259	\\
418	&	51817	&	51884	&	67	&	367	&	285	\\
419	&	51812	&	51879	&	67	&	355	&	293	\\
422	&	51811	&	51878	&	67	&	465	&	375	\\
425	&	51884	&	51898	&	14	&	370	&	290	\\
437	&	51869	&	51876	&	7	&	304	&	235	\\
440	&	51885	&	51912	&	27	&	425	&	331	\\
483	&	51902	&	51942	&	40	&	324	&	247	\\
547	&	51959	&	52207	&	248	&	347	&	288	\\
588	&	52029	&	52045	&	16	&	435	&	344	\\
594	&	52027	&	52045	&	18	&	355	&	285	\\
616	&	52374	&	52442	&	68	&	450	&	349	\\
640	&	52178	&	52200	&	22	&	405	&	323	\\
644	&	52149	&	52173	&	24	&	408	&	339	\\
662	&	52147	&	52178	&	31	&	455	&	358	\\
712	&	52179	&	52199	&	20	&	216	&	131	\\
790	&	52433	&	52441	&	8	&	402	&	323	\\
803	&	52264	&	52318	&	54	&	430	&	355	\\
804	&	52266	&	52286	&	20	&	422	&	354	\\
810	&	52326	&	52672	&	346	&	502	&	360	\\
814	&	52370	&	52443	&	73	&	363	&	279	\\
820	&	52433	&	52438	&	5	&	307	&	242	\\
960	&	52425	&	52466	&	41	&	427	&	354	\\
972	&	52428	&	52435	&	7	&	417	&	345	\\
978	&	52431	&	52441	&	10	&	328	&	262	\\
1028	&	52562	&	52884	&	322	&	466	&	371	\\
1034	&	52525	&	52813	&	288	&	471	&	385	\\
1037	&	52826	&	52878	&	52	&	508	&	417	\\
1291	&	52735	&	52738	&	3	&	421	&	337	\\
1512	&	53035	&	53742	&	707	&	144	&	99	\\
1782	&	53299	&	53383	&	84	&	457	&	373	\\
2009	&	53857	&	53904	&	47	&	449	&	354	\\
2351	&	53772	&	53786	&	14	&	332     &	259	\\
\hline 
\nodata	&	\nodata	&	\nodata	&	\nodata	&     \npairtotal\tablenotemark{e}  &  \npair\tablenotemark{e}  \\
\hline 
\label{tab:plate}
\footnotetext[1]{There are \nplate \ spectroscopic plates used, taken from the SDSS DR6.}\\
\footnotetext[2]{The plates \platesdropped \ (not listed in the table) are not used
  because the number of galaxy pairs is less than \npairminCali.}\\
\footnotetext[3]{The total number of galaxy pairs in a plate.}\\
\footnotetext[4]{The number of galaxy pairs with successful
  calibration refinement using SYS+BAND.}\\
\footnotetext[5]{The  number of galaxy pairs
  summing over all plates, respectively.}\\
\end{longtable}
\normalsize
\end{center}

\renewcommand{\thefootnote}{\arabic{footnote}}

\clearpage

\begin{table}
\begin{center}
\caption{Statistics of \fvarm \ in galaxies using different
  calibration refinements.}
\begin{tabular}{lcccc}
\hline  
method  &  mean\tablenotemark{a}    &
\sdom & \sd  \\
\hline  
             No refinement\tablenotemark{b} &	0.020	&	0.0002	&	0.025	\\
                                   SYS+BAND &	0.010	&	0.0001	&	0.009	\\
SYS+BAND (280~km~s$^{-1}$)\tablenotemark{c} &	0.012	&	0.0001	&	0.011	\\
                                   SYS+OIII &	0.016	&	0.0001	&	0.015	\\
   SYS+BAND (RED galaxies)\tablenotemark{d} &	0.010	&	0.0001	&	0.009	\\
  SYS+BAND (BLUE galaxies)\tablenotemark{d} &	0.010	&	0.0001	&	0.009	\\
\hline
\end{tabular}
\label{tab:fvar}
\tablecomments{There are  11,766 common spectral  pair in each  of the
  calibration refinement  approaches.  Each  spectrum is chosen  to be
  refined successfully in  each and every of the  above procedures. In
  principal there  can be more  objects with successful  refinement in
  each case.}  
\tablenotetext{a}{The  statistics  of  \fvarm  \  are
  calculated at spectral S/N \SNmin: \sdom \ is the standard deviation
  of  the mean  of \fvarm,  \sd \  is the  one sigma  sample scatter.}
  \tablenotetext{b}{The spectra  are the direct  output made available
  by   the   SDSS   and    no   calibration   refinement   is   done.}
  \tablenotetext{c}{Same   as  SYS+BAND   (in   which  the   rest-band
  considered      is      6450~\AA~$\pm~520$~\kms)      but      using
  6450~\AA~$\pm~280$~\kms.}   \tablenotetext{d}{Same  as SYS+BAND  but
  using  only  the  RED  or  the  BLUE  galaxies  in  calculating  the
  calibration spectrum.}
\end{center}
\end{table}

\begin{table}
\begin{center}
\caption{Upper percentile of the galaxy variable
  candidates\tablenotemark{a}.}
\begin{tabular}{lccccc}
\hline 
lower S/N  &      0 &      5 &     10 &     15 &    20 \\
upper S/N  &      5 &     10 &     15 &     20 &    40 \\
\hline
     number\tablenotemark{b} & 11	&	102	&	246	&	206	&	86	\\
 percentile\tablenotemark{c} & 1.53	&	1.89	&	3.92	&	6.26	&	3.44	\\
\hline
\end{tabular}
\label{tab:percentile}
\tablecomments{In the galaxy sample using SYS+BAND refinement.}
\tablenotetext{a}{The variable candidates are selected by \fvarm \
  $>$ (mean + \threesigma) of \fvarm \ at a given spectral S/N.}
\tablenotetext{b}{The number of variable candidates between the lower
  and upper S/N.}
\tablenotetext{c}{The percentile of the corresponding number within a
  given S/N range.}
\end{center}
\end{table}

\begin{table}
\begin{center}
\caption{Statistics of \fvarm \ in the full galaxy sample.}
\begin{tabular}{lcccc}
\hline  
   type  & number \tablenotemark{a,b} &    mean    &
   \sdom &  \sd\tablenotemark{a}  \\
\hline  
Seyfert 2	&	92	&	0.009	&	0.0008	&	0.008	\\
Starforming	&	1333	&	0.012	&	0.0003	&	0.011	\\
Composite	&	314	&	0.010	&	0.0005	&	0.009	\\
\hline
eClass A	&	729	&	0.010	&	0.0004	&	0.010	\\
eClass B	&	919	&	0.010	&	0.0003	&	0.010	\\
eClass C	&	587	&	0.011	&	0.0004	&	0.010	\\
eClass D	&	483	&	0.012	&	0.0005	&	0.010	\\
eClass E	&	3	&	0.024	&	0.0100	&	0.017	\\
eClass F	&	5	&	0.044	&	0.0124	&	0.028	\\
\hline
RED	&	9603	&	0.009	&	0.0001	&	0.009	\\
BLUE	&	2705	&	0.012	&	0.0002	&	0.011
\\
\hline
\end{tabular}
\label{tab:fvarFullType}
\tablecomments{In the galaxy sample using SYS+BAND refinement.}
\tablenotetext{a}{The number of spectra at S/N \SNmin.}
\tablenotetext{b}{We do not find any LINER \citep{1980A&A....87..152H}
  in our galaxy sample, using the classification criteria by
  \citet[][their Eqn.~15]{2006MNRAS.372..961K}.}
\end{center}
\end{table}

\begin{table}
\begin{center}
\caption{Statistics of \fvarm \ in the galaxy variable candidates.}
\begin{tabular}{lcccc}
\hline  
   type  & number \tablenotemark{a} &    mean    &
   \sdom &  \sd\tablenotemark{a}  \\
\hline 
Seyfert 2	&	5	&	0.026	&	0.0030	&
   0.007	\\
Starforming	&	62	&	0.037	&	0.0023	&
   0.018	\\
Composite	&	13	&	0.031	&	0.0043	&
   0.015	\\
\hline
eClass A	&	34	&	0.031	&	0.0027	&
   0.016	\\
eClass B	&	45	&	0.031	&	0.0020	&
   0.014	\\
eClass C	&	37	&	0.030	&	0.0015	&
   0.009	\\
eClass D	&	18	&	0.034	&	0.0030	&
   0.013	\\
\hline
RED	&	388	&	0.028	&	0.0006	&	0.012
   \\
BLUE	&	154	&	0.034	&	0.0012	&	0.015
   \\
\hline
\end{tabular}
\label{tab:fvarVariType}
\tablecomments{In the galaxy sample using SYS+BAND refinement.}
\tablenotetext{a}{The number of spectra at S/N \SNmin.}
\end{center}
\end{table}

\begin{table}
\begin{center}
{\scriptsize
\caption{Difference in AB synthetic absolute magnitude between
  the dim and bright phases in $g, r, i$ bands, for variable candidates.}
\begin{tabular}{lccccccccccccc}
\hline
type & $\left<M_g\right>$\tablenotemark{a} & 
$\left<M_r\right>$ & 
$\left<M_i\right>$ & 
$\left<z\right>$\tablenotemark{b} & 
$\left<\Delta{M_g}\right>$\tablenotemark{c} & 
$\left<\Delta{M_r}\right>$ & 
$\left<\Delta{M_i}\right>$ & 
\begin{sideways} {$\delta(\left<\Delta{M_g}\right>)$\tablenotemark{d}} \end{sideways} &
\begin{sideways} {$\delta(\left<\Delta{M_r}\right>)$} \end{sideways} & 
\begin{sideways} {$\delta(\left<\Delta{M_i}\right>)$} \end{sideways} & 
\begin{sideways} {$\sigma(\Delta{M_g})$\tablenotemark{e}} \end{sideways} & 
\begin{sideways} {$\sigma(\Delta{M_r})$}  \end{sideways} &
\begin{sideways} {$\sigma(\Delta{M_i})$} \end{sideways} \\ 
\hline
Seyfert 2	&	-20.57	&	-21.36	&	-21.30	&	0.14	&	0.07	&	0.05	&	0.05	&	0.26	&	0.12	&	0.20	&	0.01	&	0.03	&	0.03	\\
Starforming	&	-18.79	&	-19.37	&	-19.46	&	0.09	&	0.10	&	0.05	&	0.05	&	0.09	&	0.04	&	0.08	&	0.05	&	0.03	&	0.05	\\
Composite	&	-19.79	&	-20.53	&	-20.53	&	0.11	&	0.09	&	0.04	&	0.03	&	0.19	&	0.08	&	0.13	&	0.05	&	0.02	&	0.03	\\
\hline
eClass A	&	-20.70	&	-21.55	&	-21.25	&	0.15	&	0.08	&	0.05	&	0.06	&	0.11	&	0.05	&	0.85	&	0.04	&	0.04	&	0.05	\\
eClass B	&	-20.12	&	-20.95	&	-20.74	&	0.13	&	0.09	&	0.05	&	0.04	&	0.10	&	0.04	&	4.82	&	0.03	&	0.03	&	0.03	\\
eClass C	&	-19.95	&	-20.73	&	-20.81	&	0.12	&	0.09	&	0.04	&	0.04	&	0.12	&	0.05	&	0.08	&	0.03	&	0.02	&	0.03	\\
eClass D	&	-18.18	&	-18.75	&	-18.91	&	0.07	&	0.09	&	0.04	&	0.06	&	0.16	&	0.07	&	0.12	&	0.05	&	0.02	&	0.06	\\
\hline
RED	&	-20.14	&	-20.99	&	-20.91	&	0.13	&	0.08	&	0.04	&	0.05	&	0.03	&	0.01	&	3.16	&	0.04	&	0.03	&	0.04	\\
BLUE	&	-19.27	&	-19.87	&	-19.89	&	0.10	&	0.09	&	0.05	&	0.05	&	0.06	&	0.03	&	0.05	&	0.05	&	0.03	&	0.05	\\\hline
\end{tabular}
\label{tab:contvari}
\tablenotetext{a}{The sample-averaged  absolute AB
 magnitude in  the dim phase.  Unless otherwise specified,  the terms
 ``bright'' and  ``dim'' refer  to the epoch  among the two  where the
 observed spectral flux  is larger and smaller and  do not necessarily
 imply an underlying physical mechanism.}  
\tablenotetext{b}{The sample average of the redshift.} 
 \tablenotetext{c}{The sample average of the magnitude difference,
 $\Delta{M_g} = M_g(\textrm{dim}) - M_g(\textrm{bright})$,  between  the
 dim and bright phases.}  
 \tablenotetext{d}{The uncertainty  in the sample-averaged  magnitude difference,
 by  propagating  the  uncertainty   in  spectral  flux  density  to
 that in magnitude, and then to the uncertainty of the sample-averaged magnitude difference.}  
\tablenotetext{e}{The \onesigma \ sample scatter in the $\Delta{M_g}$.} 
}
\end{center}
\end{table}

\begin{table}
\begin{center}
\caption{Best-fit characteristic variability time scale.}
\begin{tabular}{lccccc}
\hline
type & zero-point \fvarm\tablenotemark{a} & $\tau_s$
(year)\tablenotemark{b} & asymptotic \fvarm (\dtaurest $\rightarrow
\infty$) & reduced \chisq & range in \dtaurest \ (day) \\
\hline
QSO 
&$     0.012\pm     0.003$&$   0.4\pm   1.0$&$     0.022\pm     0.002$&$      0.04$&
\dtauRestRangeQSO \\ 
star 
&$     0.010\pm     0.069$&$  11.2\pm   6.3$&$     0.035\pm     0.035$&$      3.41$&
\dtauRestRangeStar \\
galaxy
&$     0.009\pm     0.162$&$   8.5\pm   1.6$&$     0.025\pm     0.081$&$     19.52$&
\dtauRestRangeGALAXY \\
\hline
\end{tabular}
\label{tab:taus}
\tablenotetext{a}{\fvarm \ at zero time lag.}
\tablenotetext{b}{Characteristic variability time scale in the restframe.}
\end{center}
\end{table}

\begin{table}
\begin{center}
\caption{Best-fit characteristic variability time scale, after
  quadtrature subtraction of variability at zero time lag.}
\begin{tabular}{lcccc}
\hline
type & $\tau_s$
(year)\tablenotemark{a} & asymptotic \fvarm (\dtaurest $\rightarrow
\infty$) & reduced \chisq & range in \dtaurest \ (day) \\
\hline
QSO 
&$   0.4\pm   0.0$&$     0.018\pm     0.005$&$     13.10$&
\dtauRestRangeQSO \\ 
star 
&$   0.8\pm   0.0$&$     0.012\pm     0.003$&$     49.71$&
\dtauRestRangeStar \\
galaxy
&$   1.4\pm   0.0$&$     0.011\pm     0.018$&$    501.97$&
\dtauRestRangeGALAXY \\
\hline
\end{tabular}
\label{tab:taus_quadsubtract}
\tablenotetext{a}{Characteristic variability time scale in the restframe.}
\end{center}
\end{table}

\begin{table}
\begin{center}
\caption{BPT line-ratio difference between dim and bright phases, for variable candidates.}
\begin{tabular}{lcccc}
\hline
 type & 
\lb{$\Delta$\NIIHalpha}\rb\tablenotemark{a} & \lb{$\Delta$\OIIIHbeta}\rb &
$\sigma$($\Delta$\NIIHalpha)\tablenotemark{b} & $\sigma$($\Delta$\OIIIHbeta) \\
\hline
  Seyfert 2  &$  0.02\pm  0.03$&$  0.06\pm  0.08$&  0.03&  0.03\\
  Starforming&$  0.09\pm  0.01$&$  0.16\pm  0.03$&  0.17&  0.21\\
  Composite  &$  0.03\pm  0.03$&$  0.11\pm  0.09$&  0.03&  0.08\\
\hline
     eClass C&$  0.17\pm  0.05$&$  0.47\pm  0.12$&  0.18&  0.43\\
     eClass D&$  0.07\pm  0.02$&$  0.15\pm  0.06$&  0.08&  0.17\\
\hline
\end{tabular}
\label{tab:linevari}
\tablenotetext{a}{The sample-averaged line ratio. The quoted uncertainty is
 the    uncertainty    in     the    spectral    flux    measurement.}
 \tablenotetext{b}{The \onesigma  \ sample  scatter in the  line-ratio
 difference between dim and bright phases.}
\end{center}
\end{table}

\begin{table}
\begin{center}
\caption{Statistics of \fvarm \ in stars.}
\begin{tabular}{lcccc}
\hline  
method  &     mean    &
\sdom & \sd  \\
\hline  
                      No refinement  & 	0.035	&	0.0011	&	0.053	\\
                           SYS+BAND  & 	0.012	&	0.0003	&	0.013	\\
\hline
\end{tabular}
\tablecomments{See caption in  Table~\ref{tab:fvar} for the meaning of
  the  statistics. There  are  2,379 common spectral pairs  in  each of  the
  refinement approaches. The statistics  of \fvarm \ are calculated at
  spectral S/N \SNmin.}
\label{tab:fvar_star}
\end{center}
\end{table}

\begin{table}
\begin{center}
\caption{Statistics of \fvarm \ in QSOs.}
\begin{tabular}{lcccc}
\hline  
method  &    mean    &
\sdom & \sd  \\
\hline  
                      No refinement  & 	0.036	&	0.0012	&	0.034	\\
                           SYS+BAND  & 	0.015	&	0.0005	&	0.014	\\
\hline
\end{tabular}
\tablecomments{See caption in  Table~\ref{tab:fvar} for the meaning of
  the statistics.  There are 766  common spectral pairs in each of the
  refinement approaches. The statistics  of \fvarm \ are calculated at
  spectral S/N \SNmin.}
\label{tab:fvar_qso}
\end{center}
\end{table}

\begin{table}
\begin{center}
{\scriptsize
\caption{Matched objects between GCVS\tablenotemark{a} and our samples.}
\begin{tabular}{rrcccrrrrc}
\hline
 RA\tablenotemark{b} (degree) & DEC\tablenotemark{b} (degree) &
 specClass\tablenotemark{c} & S/N & $\Delta\tau$\tablenotemark{d} & difference\tablenotemark{e} in $g$ &
 difference in $r$ & difference in $i$ & \fvarm & nSigma\tablenotemark{f} \\  
\hline
218.750940	&	-0.768439	&	1	&	20.4	&	53	&	$	0.13	\pm	0.49	$	&	$	-0.05	\pm	0.30	$	&	$	-0.09	\pm	0.41	$	&	0.017	&	1	\\
118.540000	&	42.817851	&	1	&	39.0	&	7	&	$	-0.02	\pm	0.26	$	&	$	0.00	\pm	0.16	$	&	$	0.00	\pm	0.21	$	&	0.005	&	0	\\
119.352760	&	43.207637	&	1	&	43.0	&	7	&	$	0.05	\pm	0.23	$	&	$	0.01	\pm	0.16	$	&	$	-0.02	\pm	0.22	$	&	0.011	&	3	\\
120.915900	&	42.512477	&	1	&	46.5	&	7	&	$	-0.23	\pm	0.20	$	&	$	-0.03	\pm	0.13	$	&	$	0.08	\pm	0.18	$	&	0.051	&	4	\\
133.434070	&	57.811289	&	1	&	52.2
 &	40	&	$	0.16	\pm	0.18	$	&
 $	0.06	\pm	0.12	$	&	$	0.00	\pm
 0.18	$	&	0.045	&	4	\\
\hline  
\end{tabular}
\label{tab:gcvs}
\tablenotetext{a}{The General Catalogue of Variable Stars \citep{1999yCat.2214....0K}.} 
\tablenotetext{b}{Given in J2000.}
\tablenotetext{c}{All of the matched objects are stars according to the SDSS spectral classification.}
\tablenotetext{d}{The  time lag $\Delta\tau$  is the difference between  the two
MJDs  of   observation,  MJD(epoch  2)$-$MJD(epoch   1).}
\tablenotetext{e}{The difference in magnitude between two epochs.}
\tablenotetext{f}{The detection significance as a variable candidate, derived from this work.}
}
\end{center}
\end{table}

\begin{table}
\begin{center}
{\scriptsize
\caption{Matched objects between SDSS Stripe 82 variable catalog\tablenotemark{a} and
   stars in this work.}
\begin{tabular}{rrccccrrrrc}
\hline
\begin{sideways} {RA (degree)} \end{sideways} & 
\begin{sideways} {DEC (degree)}\end{sideways} &
\begin{sideways} {specClass} \end{sideways} & 
S/N & 
\begin{sideways} {number of obs\tablenotemark{a} in $r$} \end{sideways} & 
$\Delta\tau$ & difference in $g$ &
 difference in $r$ & difference in $i$ & \fvarm & nSigma \\  
\hline
356.032540	&	1.027191	&	1	&	18.2	&	10	&	94	&	$	0.03	\pm	1.64	$	&	$	0.01	\pm	0.45	$	&	$	0.00	\pm	0.49	$	&	0.003	&	0	\\
356.151320	&	0.303432	&	1	&	9.1	&	12	&	94	&	$	-0.01	\pm	2.13	$	&	$	-0.02	\pm	0.68	$	&	$	-0.04	\pm	0.73	$	&	0.007	&	0	\\
356.171110	&	-1.220269	&	1	&	26.8	&	10	&	94	&	$	0.08	\pm	0.53	$	&	$	0.00	\pm	0.28	$	&	$	-0.07	\pm	0.37	$	&	0.006	&	0	\\
354.635580	&	0.856453	&	1	&	7.9	&	10	&	94	&	$	0.02	\pm	1.28	$	&	$	-0.02	\pm	0.94	$	&	$	-0.09	\pm	1.68	$	&	0.039	&	1	\\
12.348661	&	-0.299891	&	1	&	52.6	&	8	&	101	&	$	0.01	\pm	0.23	$	&	$	0.00	\pm	0.15	$	&	$	0.00	\pm	0.23	$	&	0.000	&	0	\\
13.369355	&	-0.722569	&	1	&	45.5	&	8	&	101	&	$	-0.13	\pm	0.35	$	&	$	0.01	\pm	0.19	$	&	$	0.05	\pm	0.27	$	&	0.025	&	1	\\
31.675864	&	0.988745	&	1	&	17.0	&	14	&	65	&	$	-0.01	\pm	0.60	$	&	$	0.00	\pm	0.35	$	&	$	0.01	\pm	0.53	$	&	0.001	&	0	\\
36.891406	&	0.437566	&	1	&	11.7	&	14	&	421	&	$	0.03	\pm	0.79	$	&	$	0.03	\pm	0.58	$	&	$	0.02	\pm	0.97	$	&	0.013	&	0	\\
35.787734	&	0.959517	&	1	&	25.6	&	14	&	421	&	$	-0.06	\pm	0.44	$	&	$	-0.03	\pm	0.26	$	&	$	0.01	\pm	0.37	$	&	0.015	&	2	\\
36.347691	&	0.660534	&	1	&	20.4	&	13	&	421	&	$	-0.04	\pm	0.57	$	&	$	-0.03	\pm	0.33	$	&	$	-0.03	\pm	0.47	$	&	0.015	&	2	\\
44.620576	&	-0.609703	&	1	&	14.3	&	14	&	61	&	$	0.18	\pm	0.74	$	&	$	0.01	\pm	0.45	$	&	$	-0.09	\pm	0.71	$	&	0.036	&	4	\\
44.322878	&	0.785284	&	1	&	13.2	&	14	&	61	&	$	-0.14	\pm	0.73	$	&	$	-0.01	\pm	0.44	$	&	$	0.06	\pm	0.68	$	&	0.031	&	4	\\
46.057464	&	-1.220639	&	1	&	29.6	&	12	&	56	&	$	-0.16	\pm	0.32	$	&	$	-0.03	\pm	0.20	$	&	$	0.07	\pm	0.29	$	&	0.032	&	4	\\
48.387960	&	0.715234	&	1	&	17.0	&	14	&	108	&	$	-0.07	\pm	0.56	$	&	$	-0.02	\pm	0.35	$	&	$	0.03	\pm	0.58	$	&	0.014	&	1	\\
49.984166	&	-0.044137	&	1	&	8.6	&	14	&	108	&	$	-0.12	\pm	1.16	$	&	$	-0.01	\pm	0.79	$	&	$	0.05	\pm	1.39	$	&	0.030	&	0	\\
51.883714	&	0.064299	&	1	&	54.1	&	12	&	32	&	$	-0.40	\pm	0.19	$	&	$	-0.05	\pm	0.12	$	&	$	0.15	\pm	0.17	$	&	0.085	&	4	\\
52.295747	&	0.603985	&	1	&	20.9	&	12	&	32	&	$	-0.08	\pm	0.62	$	&	$	-0.01	\pm	0.34	$	&	$	0.04	\pm	0.49	$	&	0.015	&	2	\\
52.476904	&	0.389118	&	1	&	45.5	&	14	&	32	&	$	0.02	\pm	0.28	$	&	$	0.00	\pm	0.17	$	&	$	0.00	\pm	0.24	$	&	0.006	&	0	\\
51.602434	&	0.674580	&	1	&	45.2	&	14	&	20	&	$	0.01	\pm	0.24	$	&	$	0.01	\pm	0.16	$	&	$	-0.04	\pm	0.21	$	&	0.007	&	4	\\
43.924804	&	-0.542138	&	1	&	2.8	&	14	&	707	&	$	-0.10	\pm	4.41	$	&	$	0.15	\pm	2.26	$	&	$	0.17	\pm	3.21	$	&	0.017	&	0	\\
44.201206	&	0.013831	&	1	&	19.2	&	14	&	707	&	$	-0.15	\pm	1.10	$	&	$	-0.01	\pm	0.30	$	&	$	0.03	\pm	0.30	$	&	0.011	&	0	\\
44.918114	&	0.758511	&	1	&	6.3
 &	12	&	707	&	$	-0.14	\pm	2.89  $	&	$	-0.17	\pm	0.98	$	&	$ -0.28	\pm	1.13	$	&	0.100	&	2	\\ 
\hline  
\end{tabular}
\label{tab:varStripe82_star}
\tablecomments{See also captions in Table~\ref{tab:gcvs}.}
\tablenotetext{a}{Given by \citet{2007AJ....134.2236S}.} 
}
\end{center}
\end{table}

\begin{table}
\begin{center}
\caption{Matched objects between SDSS Stripe 82 variable catalog\tablenotemark{a} and
  QSOs in this work}
\begin{tabular}{ccc}
\hline  
nSigma\tablenotemark{b} & number of objects & percentage \\
\hline  
0	&	231	&	67.3	\\
1	&	45	&	13.1	\\
2	&	33	&	9.6	\\
3	&	10	&	2.9	\\
4	&	24	&	7.0	\\
\hline
\end{tabular}
\label{tab:varStripe82_qso}
\tablenotetext{a}{Given by \citet{2007AJ....134.2236S}.} 
\tablenotetext{b}{The detection significance as a variable candidate, derived from
  this work.}
\end{center}
\end{table}

\end{document}